\documentclass[10pt]{book}
\usepackage{array}

\def\Null{\mathop{\rm Null}}
\def\Range{\mathop{\rm Range}}
\usepackage{amssymb,latexsym,amsmath, graphicx, amsfonts, natbib, ulem}
\usepackage{color}

\begin{document}


\renewcommand{\baselinestretch}{1.2}

\markright{
\hbox{\footnotesize\rm Statistica Sinica (2013): Preprint}\hfill
}

\markboth{\hfill{\footnotesize\rm MADAN KUNDU, JAROSLAW HAREZLAK AND TIMOTHY  RANDOLPH} \hfill}
{\hfill {\footnotesize\rm LONGITUDINAL FUNCTIONAL MODELS WITH STRUCTURED PENALTIES} \hfill}

\renewcommand{\thefootnote}{}
$\ $\par


\fontsize{10.95}{14pt plus.8pt minus .6pt}\selectfont
\vspace{0.8pc}
\centerline{\large\bf Longitudinal Functional Models with Structured Penalties}
\vspace{.4cm}
\centerline{Madan G. Kundu$^1$, Jaroslaw Harezlak$^1$ and Timothy W. Randolph$^2$}
\vspace{.4cm}
\centerline{\it $^1$Department of Biostatistics}
\centerline{\it Indiana University Fairbanks School of Public Health}
\centerline{\it Indianapolis, USA}
\vspace{.4cm}
\centerline{\it $^2$Biostatistics and Biomathematics Program}
\centerline{\it Fred Hutchinson Cancer Research Center}
\centerline{\it Seattle, USA}
\vspace{.55cm}
\fontsize{9}{11.5pt plus.8pt minus .6pt}\selectfont


\begin{quotation}
\noindent {\it Abstract:} This paper addresses estimation in a longitudinal regression
model for association between a scalar outcome and a set of longitudinally-collected
functional covariates or predictor curves.  The framework consists of estimating a
time-varying coefficient function that is modeled as a linear combination of
time-invariant functions but having time-varying coefficients.  The estimation procedure
exploits the equivalence between penalized least squares estimation and a linear mixed
model representation. The process is empirically evaluated with several simulations and
it is applied to analyze the neurocognitive impairment of HIV patients and its
association with longitudinally-collected magnetic resonance spectroscopy curves.\par

\vspace{9pt} \noindent {\it Key words and phrases:} Functional data analysis,
longitudinal data, mixed model, structured penalty, generalized singular value
decomposition
\par
\end{quotation}\par


\fontsize{10.95}{14pt plus.8pt minus .6pt}\selectfont

\section{Introduction}\label{Intro}
Technological advancements and increased availability of storage of large datasets have
allowed for the collection of functional data as part of time-course or longitudinal
studies.  In the cross-sectional setting, there have been many proposed methods for
estimating a regression function in a so-called functional linear model (fLM).  This
function is a functional (continuous) analogue of a vector of (discrete) regression
coefficients; it connects the scalar response, $y$ to a functional covariate, $w\equiv
w(s)$.  Although these models have recently been well studied, extensions to
longitudinally-collected functions have not received much attention. Only recently
longitudinal penalized functional regression (LPFR) and longitudinal functional principal
component regression (LFPCR) approaches have been proposed to extend the cross-sectional
fLM to a longitudinal setting by incorporating subject-specific random intercepts
\citep{LPFR, gertheiss2013longitudinal}. A basic assumption in both LPFR and LFPCR is
that the regression function remains constant over time. Consequently, these methods are
not suited for situations in which the association between a functional predictor and
scalar response may evolve over time. Here we propose a technique that extends the
analysis of functional linear models by relating a scalar outcome to a functional
predictor---both observed longitudinally---and estimates a time-dependent regression
function.

The method fits into a generalized ridge regression framework by imposing a
scientifically-informed quadratic penalty term into the estimation process.
The extension of this framework to the longitudinal
setting has two major advantages: 1) the regression function is allowed to vary over
time; and 2) external or a priori information about the structure of the regression
function can be incorporated directly into the estimation process. We formulate the
estimation procedure within a mixed-model framework making the method computationally
efficient and easy to implement.

\citet{FDATools} introduced the term functional data analysis (FDA) in the statistical
literature. The cross-sectional fLM with scalar response can be stated as follows
\citep[see e.g., ][]{QuadFun}
\begin{equation*}
E(y|W)=\mu_y +\int_{\Omega}{W(s) \gamma(s) ds}
\end{equation*}
where $\mu_y$ is the mean of $y$, $\Omega$ denotes the domain of the predictor functions
$W(s)$, $s\in\Omega$, and $\gamma(s)$ is a square integrable function that models the
linear relationship between the functional predictor and scalar response.  We will assume
that $W(\cdot)$ denotes a mean-centered function ($E[W(s)] = 0$ for almost all $s\in
\Omega$).

As there is no unique $\gamma(\cdot)$ that solves this equation some form of
regularization, or constraint, is required. For example, a common approach is to impose
smoothness on $\gamma(\cdot)$. One approach to this is to expand both the regression
function $\gamma(\cdot)$ and predictor functions $W(\cdot)$ in terms of B-splines and
then obtain the regularized estimate of $\gamma(\cdot)$ \citep{FDABook}. Another approach
is to express the regression function $\gamma(\cdot)$ in terms of the empirical
orthonormal basis obtained by the eigenfunctions of the covariance of $W(\cdot)$ (i.e., a
Karhunen-Lo\`eve (K-L) expansion \citep[see e.g., ][]{Muller2005}). A third approach,
known as penalized functional regression (PFR) \citep{PFR}, combines the above two
methods. In PFR, a spline basis is used to represent $\gamma(\cdot)$ and a subset of
empirical eigenfunctions is used to represent each $W(\cdot)$.  Another approach is to
use a wavelet basis, instead of splines or eigenfunctions, to represent the predictor
functions \citep{MorrisCarroll}.

Here we adopt an approach by \citet{PEER} which does not begin by explicitly projecting
onto a pre-specified basis of functions.  Instead, prior information about functional
structure is incorporated into the estimation process by way of a penalty operator, $L$.
This approach of ``partially empirical eigenvectors for regression" (PEER) exploits the
fact that a penalized least-squares regression estimate mathematically arises as a series
expansion in terms of a set of basis functions determined {\it jointly} by the covariance
(empirical functional structure) and the penalty (imposed structure); see also the
Appendix~\ref{GSVD}.  This naturally extends ridge regression (non-stuctured penalty) and
smoothing penalties such as a second-derivative penalty (presuming a smooth regression
function). Here we extend the scope of the PEER approach to the longitudinal setting in a
manner that allows the estimated regression function $\gamma\equiv\gamma(t,\cdot)$ to
vary with time.

An important concern for any regularization method is identifiability of the estimate;
i.e., the lack of uniqueness or, possibly, its instability. In FDA this arises from the
lack of invertibility of the empirical covariance operator: a finite number of predictor
curves means the dimension of the range of this operator is finite and so, as an operator
on a infinite-dimensional domain, it has a non-trivial null space.  The philosophy behind
a penalty-operator approach is that estimation is constrained to the subspace spanned by
functions that are the jointly determined by $W$ and $L$.  A sufficient condition for
uniqueness of this estimate is to assume $\Null(W)\cap \Null(L)=\{0\}$; see \citep{Engl}
or \citep{Bjorck}. We assume this throughout.

\begin{figure}
\centerline{%
\includegraphics [angle=0,width=135mm, height=80mm]{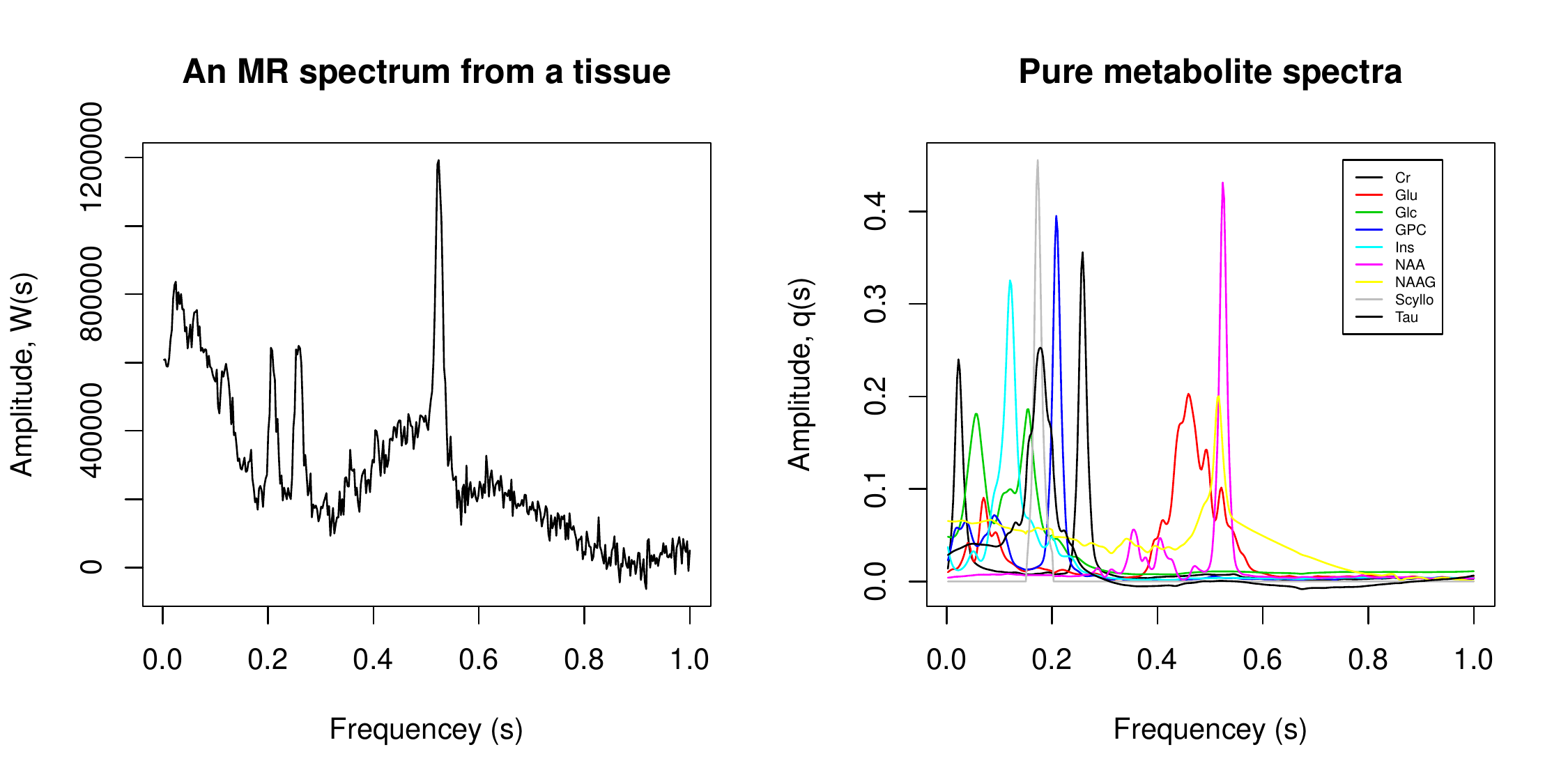}}
\caption{{\it Left panel}: an observed MR spectrum from tissue.  {\it Right panel}: The 9 pure metabolite spectra.
In each plot, the $y$-axis represents amplitude and $x$-axis the frequency of nucleus, $s$, transformed to $[0,1]$ interval.}
\label{fig:App1}
\end{figure}

The problem we address involves repeated observations from each of $N$ subjects. For each
subject, $i$, at each observation time, $t$, we collect data on a scalar response
variable, $y$, and a (idealized) predictor function, $W(\cdot)$. We are interested in
longitudinal regression models of the form:
\begin{equation} \label{eq:first}
y_{it}=x_{it}^{\top} \beta +\int_0^1{W_{it}(s) \gamma (t,s) ds} +z_{it}^{\top}b_i +\epsilon_{it}.
\end{equation}
Here $\gamma (t,\cdot)$ denotes the regression function at time $t$, $x_{it}$ is a vector
of scalar-valued (non-functional) predictors; $z_{it}^{\top}b_i$ and $\epsilon_{it}$
denote the subject specific random effect and random error term, respectively. In a
spirit similar to that of a linear mixed model with time-related slope for longitudinal
data, we assume that $\gamma(t,\cdot)$ can be decomposed into several time-invariant
component functions; e.g., $\gamma(t,\cdot) =\gamma_0(\cdot)+ t\;\gamma_1(\cdot)$.

Our work is motivated by a study in which magnetic resonance (MR) spectra have been
collected longitudinally from late stage HIV patients \citep{Harezlak2011}. We consider
global deficit score (GDS) as a scalar response variable, $y$, and MR spectra as
predictor functions, $W(\cdot)$. Of interest is the association of GDS with MR spectra
and how this association evolves with time.
One MR spectrum is shown in the left panel of Figure~\ref{fig:App1}: the amplitude,
$W(s)$, is plotted against the transformed frequency of nucleus, $s$, to the $[0,1]$ interval ($x$-axis).
The pattern and amplitudes of the peaks contain information about the concentration of metabolites
present in tissue. Each metabolite has a {\it unique} spectrum and so one MR spectrum is
a mixture of spectra from each individual metabolite (plus background and random noise);
see the right panel in Figure~\ref{fig:App1} which displays spectra from 9 metabolites.
Consequently, one expects an observed spectrum from tissue to lie near a functional
subspace, $\mathcal{Q}$, spanned by the spectra of pure metabolites. The regression
function, $\gamma(t, \cdot)$, models the association between $y$ and $W(\cdot)$ and
hence, in principle, should also lie near $\mathcal{Q}$. Hence, the subspace
$\mathcal{Q}$ should be more informative than B-splines or cosine functions that are in
some sense``external" to the problem. For this reason, we adopt a methodology that
encourages the estimate of $\gamma(\cdot)$ to be near to $\mathcal{Q}$.  The approach is
implemented using a {\it decomposition based penalty} which penalizes the estimate of
$\gamma(t, \cdot)$ lightly if it belongs to $\mathcal{Q}$ and strongly if it does not
\citep{PEER}.

The cross-sectional fLM with scalar response has been a focus of various investigations
\citep{FDABook, Faraway1997, Fan2000, cardot1999, cardot2003, Cai2006, cardot2007,
Reiss2009}, many of which estimate a regression function in two steps.  For example,
\citet{cardot2003} first perform principal component regression (PCR), which projects the
observed predictor curves onto an empirical basis to obtain an estimate, then use
B-splines to smooth the result. \citet{Reiss2009} study several of these methods along
with modifications that include versions of PCR using B-splines and second-derivative
penalties (cf. \citep{FDABook,Silverman1996}). Extensions of fLM have been made towards
generalized linear model with functional predictors
\citep{james2002generalized,Muller2005generalized} and quadratic functional regression
\citep{QuadFun}. We are interested in extending the fLM to a longitudinal setting.

To our knowledge, the only published methods addressing the longitudinal functional
predictor framework are LPFR \citep{LPFR} and LFPCR \citep{gertheiss2013longitudinal}.
The  LPFR approach assumes the regression function in \eqref{eq:first} is independent of
time and proceeds in three steps: use a truncated set of K-L vectors to represent the
predictor functions; express the regression function with a spline basis; fit the
longitudinal model using an equivalent mixed-model framework that incorporates
subject-specific random effects.  In the LFPCR approach, the predictor functions are
first decomposed into visit- and subject-specific functions accordingly via longitudinal
functional principal component analysis (LFPCA) \citep{Greven2011} and in a second step,
longitudinal analysis is carried out with the outcome of LFPCA.   Both LPFR and LFPCR
assume that the regression function, $\gamma(t, \cdot)$ remains constant over time. In
contrast, we model the coefficient function $\gamma(t,\cdot)$ as a time-dependent linear
combination of several time-invariant component functions, $\{\gamma_d(\cdot)\}_{d=0}^D$,
each of which is estimated via a penalty operator that is informed by the structure of
the data or a scientific question.

Section~\ref{Model} establishes notation for the model considered in this paper. In
Section~\ref{Ridge}, the concept of generalized ridge \citep{HoerlKen1970} (or
\citet{Tikhonov}) estimation is discussed.  We review a decomposition-based penalty in
Section~\ref{Decomp} and present how these estimates can be obtained as best linear
unbiased predictors (BLUP) through mixed model equivalence in Section \ref{BLUP}.
Expressions for the precision of the estimates are derived in Section~\ref{Precision}. In
an Appendix (Section~\ref{GSVD}) we present how our longitudinal penalized estimate,
along with its bias and precision, can be obtained, under some weak assumptions, in terms
of generalized singular vectors.

Numerical illustrations are provided by simulations in Section~\ref{Simulation}:
Section~\ref{Simulation1} compares LPFR with the method proposed in this paper;
Section~\ref{Simulation2} evaluates the influence of sample size and the effect of using
prior functional information; Section~\ref{Simulation3} explores confidence band coverage
probabilities; Section~\ref{Simulation4} evaluates performance when only partial
information is available. An application to real MRS data using and a summary of our
findings is presented in Section \ref{Application}.  The methods discussed in this paper
have been implemented in the \verb|R| package \verb|refund| \citep{refund} via the
\verb|peer()| and \verb|lpeer()| functions.

\section{Statistical Model}\label{Model}
We consider $\Omega=[0,1]$, a closed interval in $\mathbb{R}$, and let $W(\cdot)$ denotes
a random function in $L^2(\Omega)$.  Let $W_{it}(\cdot)$ denotes a  predictor function
from the $i^{th}$ subject ($i=1,\dots,N$) at the $t^{th}$ timepoint ($t =
t_1,\dots,t_{n_i}$). Technically, an observed predictor arises as a discretized sampling
from an idealized function, and we will assume that each observed predictor is sampled at
the same $p$ locations, $s_1,\dots,s_p \in [0,1]$, with sampling that is appropriately
regular and dense enough to capture informative functional structure, as seen, for
instance, in the MRS data in Section~\ref{Application}. Let
$w_{it}:=[w_{it}(s_1),\cdots,w_{it}(s_p)]^{\top}$ be the $p\times 1$ vector of values
sampled from the realized function $W_{it}(\cdot)$. Then, the observed data are of the
form $\{y_{it}; x_{it}; w_{it}\}$, where $y_{it}$ is a scalar outcome, $x_{it}$ is a
$K\times 1$ column vector of measurements on $K$ scalar predictors, and $w_{it}$ is the
sampled predictor from the $i^{th}$ subject at time $t$. Denoting the true regression
function at time $t$ by $\gamma (t,\cdot)$, the longitudinal functional regression
outcome model of interest is
\begin{equation}\label{eq:longflm}
y_{it}=x_{it}^{\top} \beta +\int_0^1{W_{it}(s) \gamma (t,s) ds} +z_{it}^{\top}b_i +\epsilon_{it}
\end{equation}
where, $\epsilon_{it} \sim  N(0,\sigma_{\epsilon}^2)$ and $b_i$ is the vector of $r$
random effects pertaining to subject $i$ and distributed as $N(0, \Sigma_{b_i})$. As
usual we assume that $z_{it}$ is a subset of $x_{it}$, $\epsilon_{it}$ and $b_i$ are
independent, $\epsilon_{it}$ and $\epsilon_{i't'}$ are independent whenever $i \neq i'$
or $t \neq t'$ or both, and $b_i$ and $b_{i'}$ are independent if $i \neq i'$. Here
$x_{it}^{\top}\beta$ is the standard fixed effect  from $K$ univariate predictors,
$z_{it}^{\top}b_i$ is the standard random effect and $\int_0^1{W_{it}(s)\gamma (t,s) ds}$
is the subject/time specific functional effect. We assume that $\gamma (t,\cdot) \in
L^2(\Omega)$, for all $t$.

The functional structure, indexed by $s$, and time structure, indexed by $t$, have
somewhat unequal roles in our model, as we assume the longitudinal observations are more
limited in the amount of information relative to the densely-sampled $s$ index. For
example, $\gamma(t,s)$ may vary linearly with time, $\gamma(t,s)=\gamma _0(s)+t\gamma
_1(s)$, or quadratically, $\gamma(t,s)=\gamma _0(s)+t\gamma_1(s)+t^2\gamma_2(s)$. This is
similar in spirit to a linear mixed effects model with linear or quadratic time slope
\citep[see e.g., ][]{ALA}. In general, we assume that $\gamma(t,s)$ can be decomposed
into several time-invariant component functions $\gamma _0(s), \cdots, \gamma_D(s)$ as
\begin{equation*}
\gamma(t,s)=\gamma_0(s)+f_1(t) \gamma_1(s)+ \ldots +f_D(t) \gamma_D(s)
\end{equation*}
where, $f_1,\dots, f_D$ are $D$ prescribed linearly independent functions of $t$ and
$f_d(0)=0$ for all $d$; the time component $t$ enters into $\gamma(t,s)$ through these
terms. At $t=0$, $\gamma(t,s)$ reduces to $\gamma_0(s)$ and has the obvious
interpretation of a baseline regression function pertaining to the sampling points $s$.
When $D=0$, $\gamma(t,s)\equiv\gamma_0(s)$ is independent of $t$, a situation considered
by \citet{LPFR}. In general, each $f$ may be any function of $t$ with $f(0)=0$, e.g.,
$f(t)=t$ or  $t\,\mbox{exp}(t)$. We can rewrite the equation~\eqref{eq:longflm} as
\begin{equation*}\label{eq:longflm-f}
y_{it}=x_{it}^{\top} \beta +\int_0^1{W_{it}(s) \{\gamma _0(s)+f_1(t) \gamma _1(s)+ \ldots + f_D(t).
   \gamma_D(s)\} ds} +z_{it}^{\top}b_i+\epsilon_{it}
\end{equation*}
The association of $y_{it}$ with $W_{it}$ is modeled as a linear dependence on
observations at $p$ sampling points, $w_{it}$.  In our approach, the (functional)
structure is imposed directly into the estimation of each $\gamma_d=[\gamma_d(s_1),
\dots,\gamma_d(s_p)]^{\top}$, for $d=0,\dots,D$ (as described in
Section~\ref{Estimation}). Combining all $n_{\bullet}=\sum_{i=1}^N{n_i}$ observations
from the $N$ subjects obtained across all time points, we express the model as
\begin{equation}
y=X \beta + W \gamma +Zb+\epsilon.
\end{equation}
Here, $y=[y_{1t_1}, \cdots , y_{1t_{n_1}}, \dots, y_{1t_N}, \dots , y_{Nt_{n_N}}]^{\top}$
is a $n_{\bullet} \times 1$ vector of all responses, $X=[x^{\top}_{1t_1}, \cdots ,$ $
x^{\top}_{1t_{n_1}}, \cdots, x^{\top}_{1t_N}, \cdots , x^{\top}_{Nt_{n_N}}]^{\top}$ is an
$n_{\bullet} \times K$ design matrix pertaining to $K$ univariate predictors, $\beta$ is
the associated coefficient vector, $\gamma=[\gamma^{\top}_0, \gamma^{\top}_1, \cdots,
\gamma^{\top}_D]^{\top}$ is a $(D+1)p \times 1$ vector of functional coefficients, $W$ is
the corresponding $n_{\bullet} \times (D+1)p$ design matrix. Further, $b$ is the $rN
\times 1$ vector of random effects and $Z$ is the corresponding $n_{\bullet} \times rN$
design matrix. The matrix $W$ has the structure
\[
W=\left[
\begin{matrix}
W_1\\\vdots \\W_N
\end{matrix}
\right]
\qquad
W_i=
\left[
\begin{matrix}
w_{it_1}^{\top} &  f_1(t_1) w_{it_1}^{\top}  & \cdots  &   f_D(t_1) w_{it_1}^{\top}  \\
 \vdots         & \vdots &     \ddots    & \vdots    \\
w_{it_{n_i}}^{\top} &  f_1(t_{n_i}) w_{it_{n_i}}^{\top}  & \cdots  &   f_D(t_{n_i}) w_{it_{n_i}}^{\top}  \\
\end{matrix}
\right]
\]

\section{Estimation of Parameters with a Penalty}\label{Estimation}
Our approach builds on intuition from single-level functional regression that encourages
an estimate of $\gamma(\cdot)$ to be in or near a ``preferred'' space via choice of
penalty operator \citep{PEER}. To describe the effect of a general penalty operator, $L$,
it is useful to consider the familiar example of a Laplacian penalty, $\mathcal{L}$.  The
typical heuristic for this arises by viewing $\beta$ as a function whose local
``smoothness" is informative. In this case, the term $||\mathcal{L}\beta||^2$ penalizes
sharp changes in $\beta$. For our perspective, it is helpful to recall that the dominant
eigenvectors of $\mathcal{L}$ (those corresponding to the largest eigenvalues) are
sharply oscillatory while the least-dominant eigenvectors are very smooth.  Hence a
linear-algebraic view of this is that rather than penalizing sharp changes, smoothness in
the estimate is inherited from the eigenproperties of $\mathcal{L}$.  More specifically,
structure in the estimate arises from the joint eigenproperties of $X$ and $\mathcal{L}$
(as given by the GSVD). In general, the least-dominant eigenvectors of a penalty $L$ will
have the largest effect on the estimate.  This property can be used to construct a
``preferred subspace" by defining a penalty $L$ whose least-dominant (or perhaps
zero-associated) eigenvectors are preferred. The steps in PEER approach are as follows:
(1) Identify the functional space where $W(\cdot)$ is expected to belong and treat this
as a ``preferred'' space; (2) define a {\it decomposition-based penalty} (see
Section~\ref{Decomp}) that penalizes more when the estimate of $\gamma$ falls into the
non-preferred space compared to preferred space. (3) Estimate $\gamma(\cdot)$ as a
penalized estimate. In our longitudinal setting, we encourage the estimates for each of
the $\gamma_0(\cdot), \cdots , \gamma_D(\cdot)$ to be close to a preferred functional
subspace.  Our estimation approach allows the preferred subspace to be different for each
of the $\gamma_d(\cdot)$'s. In the longitudinal (or $t$) dimension, $\gamma$ is more
explicitly and severely constrained by the choice of $f_1,\dots,f_D$.

\subsection{Generalized Ridge Estimate} \label{Ridge} The model described in
the previous section can be written as
\begin{equation}
y=X \beta + W \gamma+\epsilon^*,
\end{equation}
where $\epsilon^*=Zb+\epsilon \sim  N(0,V)$ and
$V=Z\Sigma_bZ^{\top}+\sigma_{\epsilon}^2I$.  Fore each $d=0,\ldots,D$, let $L_d$ be the
penalty operator for $\gamma_d$ and let $\lambda^2_d$  be the associated tuning
parameter. The corresponding penalized estimates of $\beta$ and $\gamma$ are minimizers
of:
\begin{equation} \label{minimize}
||y-X \beta - W \gamma ||_{V^{-1}}^2 + \lambda_0^2 ||\gamma_0 ||^2_{L_0^{\top}L_0}+\cdots + \lambda_D^2 ||\gamma_D ||^2_{L_D^{\top}L_D}.
\end{equation}
Here we use the notation $||a||_B^2=a^{\top}Ba$, where $B$ is a symmetric, positive
definite matrix. A generalized ridge estimate of $\beta$ and $\gamma$ based on minimizing
the above expression is obtained as  \citep[see e.g., ][p.~66]{spreg}
\begin{equation}
\left[
\begin{matrix}
\hat{\beta} \\ \hat{\gamma} \\
\end{matrix}
\right]
=(C^{\top}V^{-1}C+ D)^{-1}C^{\top}V^{-1}y           \label{RidgeEst}
\end{equation}
where, $C=[X \; \; W]$, $D=\mbox{blockdiag}\{0, L^{\top}L\}$ and
$L=\mbox{blockdiag}\{\lambda_0L_0, \cdots, \lambda_DL_D\}$.

In the Appendix, we derive an expression for the generalized ridge estimate
$\hat{\gamma}$ explicitly in terms of the generalized singular value decomposition (GSVD)
components.

\subsection{Decomposition based penalty}\label{Decomp}
Let $\gamma_d\equiv \gamma_{L_d,\lambda_d}$ be the estimate obtained from the penalty
operator $L_d$ and tuning parameter $\lambda^2_d$, for each $d=0, \ldots, D$. For
example, $L_d$ may denote $I_p$ (a ridge penalty) or a second-order derivative penalty
(giving an estimate having continuous second derivative). Alternatively, with prior
knowledge about potentially relevant structure in a regression function, a targeted
decomposition-based penalty can be defined in terms of a subspace defined by such
structure \citep{PEER}.  To be precise, if it is appropriate to impose
scientifically-informed constraints on the ``signal" being estimated by $\gamma$, this
prior may be implemented by encouraging the estimate to be in or near a subspace,
$\mathcal{Q} \subset L^2(\Omega)$.

Returning to our notation that reflects functional predictors observed at $p$ sampling
points, we represent $\mathcal{Q}$ by the range of a $p \times J$ matrix $Q$ whose
columns are $q_1,\ldots,q_J$.  Consider the orthogonal projection  $P_Q=QQ^+$ onto the
$\Range(Q)$, where $Q^+$ is Moore-Penrose inverse of $Q$. Then a decomposition penalty is
defined as
\begin{equation}
L_Q = \phi_bP_Q+\phi_a(I-P_Q) \label{decomp}
\end{equation}
for scalars $\phi_a$ and $\phi_b$. To see how $L_Q$ works, let $\tilde{\gamma}_d$ be any
estimate of $\gamma_d$. When $\tilde{\gamma}_d \in \operatorname{span}(Q)$, we have
$L_Q\tilde{\gamma}_d=\phi_b\tilde{\gamma}_d$, but when $\tilde{\gamma}_d \notin
\operatorname{span}(Q)$, we have $L_Q\tilde{\gamma}_d=\phi_a\tilde{\gamma}_d$.  The
condition $\phi_a>\phi_b$ imposes more penalty for $\tilde{\gamma}_d \notin
\operatorname{span}(Q)$ compared to when $\tilde{\gamma}_d \in \operatorname{span}(Q)$.
The weights $\phi_a$ and $\phi_b$ determine the relative strength of emphasizing
$\mathcal{Q}$ in the estimation process.  Note, in particular, that taking
$\phi_a=\phi_b$ results in a ridge estimate and that $L_Q$ is invertible, provided
$\phi_a$ and $\phi_b$ are nonzero.  Some analytical properties for this family of
penalized estimates are discussed in \citet{PEER}.

\section{ Mixed model representation}\label{Mixed}
Estimates of $\beta$ and $\gamma$ obtained by minimizing the expression in
equation~\eqref{minimize} correspond to a generalized ridge estimate. In this section we
aim to construct an appropriate mixed model that minimizes the expression in
equation~\eqref{minimize}. In general, the penalty, $L$, is not required to be invertible
but for simplicity this will be assumed here. The mixed model approach provides an
automatic selection of tuning parameters $\lambda_1, \cdots, \lambda_D$. REML-based
estimation of the tuning parameters has been shown to perform as well as the other
criteria and under certain conditions it is less variable than GCV-based estimation
\citep{Reiss2009}.

\subsection{Estimation of parameters} \label{BLUP}
Using Henderson's justification \citep{Henderson1950}, one can show that, for each
$d=0,\ldots,D$, the model $y=X\beta + W \gamma + \epsilon^*$  where, $ \epsilon^* \sim
N(0,V)$ and $\gamma_d \sim N(0, \frac{1}{\lambda_d^2}(L_d^{\top}L_d)^{-1})$, minimizes
the expression in equation~\eqref{minimize} to obtain the BLUP.  Thus the generalized
ridge estimate of $\beta$ and $\gamma$ correspond to the BLUP from the following model:
\begin{equation*}
y=X\beta + W^* \gamma^* +\epsilon
\end{equation*}
where, $W^*=[W \mbox{  } Z]$,  $\gamma^*=[\gamma^{\top} \mbox{  } b^{\top}]^{\top} \sim N[0,
\Sigma_{\gamma^*}]$ and $\epsilon \sim  N(0,\sigma_{\epsilon}^2 I)$
with
\[
 \Sigma_{\gamma^*}=\mbox{blockdiag}\{(L^{\top}L)^{-1} ,\;\;\;   \Sigma_b\}
\quad \mbox{and} \quad \Sigma_b=\mbox{blockdiag}\{\Sigma_{b_1}, \cdots , \Sigma_{b_N} \}.
\]
This representation allows us to estimate fixed and functional predictors simply by
fitting a linear  mixed model (e.g., using the \verb|lme()| of the \verb|nlme| package in
\verb|R| or \verb|PROC MIXED| in \verb|SAS|).

\subsection{Precision of Estimates}\label{Precision}
Our ridge estimate is the BLUP from an equivalent mixed model, hence the variance of the
estimate depends on whether the parameters are random or fixed. Randomness of $\gamma$ is
a device used to obtain the ridge estimate while $\epsilon$ and $b$ in our case are truly
random.
With this in mind, we follow \citet{spreg} and assume that the variance of the
estimates is conditional on $\gamma$, but not on $b$. The BLUP of $\beta$, $\gamma$ and
$b$ can be expressed as \citep[see e.g.,][]{BLUP, spreg}:
\[
\tilde{\beta} = \left(X^{\top}V_1^{-1}X\right)^{-1}X^{\top}V_1^{-1}y
\qquad
\tilde{\gamma} =( L^{\top}L)^{-1}W^{\top}V_1^{-1}(y-X\tilde{\beta})
\]
\[
\tilde{b} = \Sigma_bZ^{\top}V_1^{-1}(y-X\tilde{\beta})
\]
where $V_1=V+W( L^{\top}L)^{-1}W^{\top}$. $\tilde{\beta}$ is an unbiased estimator of $\beta$, but
$\tilde{\gamma}$ is not unbiased. It is trivial to see that $\mbox{Cov}(y|\gamma)=V$.
Thus, the variances of $\tilde{\beta}$ and $\tilde{\gamma}$, conditional on $\gamma$,
are:
\begin{equation*}
\mbox{Cov}(\tilde{\beta}|\gamma) = \left(X^{\top}V_1^{-1}X\right)^{-1}X^{\top}V_1^{-1}VV_1^{-1}X\left(X^{\top}V_1^{-1}X\right)^{-1}
\end{equation*}
\begin{equation}
\mbox{Cov}(\tilde{\gamma}|\gamma) =A_{\gamma}VA_{\gamma}^{\top}
\qquad A_{\gamma}=( L^{\top}L)^{-1}W^{\top}V_1^{-1}\{V_1-X(X^{\top}V_1^{-1}X)^{\top}\}V_1^{-1} \label{gammase}
\end{equation}

To obtain the unconditional variance, one must replace $V$ by $V_1$ in the above
expressions, but this will overestimate the variance of the estimates. Expressions for
the predicted value of $y$  and its variance are:
\begin{equation*}
\tilde{y} = X \tilde{\beta} + W \tilde{\gamma} +Z\tilde{b}
\qquad
\mbox{Cov}(\tilde{y}|\gamma) = A_yVA_y^{\top}
\end{equation*}
where $
A_y=[\{V_1-WL^{\top}LW-Z\Sigma_bZ^{\top}\}^{-1}X\left(X^{\top}V^{-1}X\right)^{-1}X^{\top}V^{-1}
+ W L^{\top}LW^{\top} + Z \Sigma_bZ^{\top}]V_1^{-1}.$

Let, $T=[1 \mbox{  }f_1(t)  \mbox{  }\cdots \mbox{  }f_d(t)]\otimes I_K$ . Then the
discretized version of regression function at time $t$ is  $\gamma_{(t)}=[\gamma(t, s_1),
\cdots, \gamma(t, s_K)] = T\gamma$. Therefore, the estimate of $\gamma_{(t)}$ is
$\tilde{\gamma}_{(t)}=T\tilde{\gamma}$ and the estimate of its variance is
$T\mbox{Cov}(\tilde{\gamma}|\gamma) T^{\top}$.
The smoothing parameters' estimates are the ratios of the variance components in the mixed model equivalence of the LongPEER model.
The derivations above do assume the knowledge of the variance components' true values. In practice,
these variance components are estimated and the empirical versions of the regression parameters are obtained (EBLUPs).

\subsection{Selection of time-structure in $\gamma(t, \cdot)$} \label{TimeStrSel}
The proposed approach allows a flexible choice of the time structure to be included in
the regression function $\gamma(t, \cdot)$. In practice, data and information to estimate
structure of the longitudinal observations (along the $t$ index) are more limited than
the functional relationship along the $s$ index. For example, whether $\gamma_0(t,
\cdot)+ t \gamma_1(t, \cdot)$ is sufficient or the more flexible $\gamma_0(t, \cdot)+ t
\gamma_1(t, \cdot)+ t^2 \gamma_2(t, \cdot)$ is required is not known. The problem of
choosing appropriate time-structure in $\gamma(t, \cdot)$
is similar, in principle, to that of choosing time structure in a linear mixed-effects
model (e.g., $E(y_{it}|b_i)=\beta_0 + \beta_1\; t$ or $E(y_{it}|b_i)=\beta_0 + \beta_1\;
t+ \beta_2\; t^2$).
We propose two approaches to decide what the form of unknown regression function is: (a) Use of the AIC to compare different structures, and (b) Use of the point-wise confidence band for the component functions: $\gamma_0(s), \ldots, \gamma_D(s)$. If the confidence band for any $\gamma_d(s)$ contains zero in its entire domain, then such term is dropped from the $\gamma(t,s)$.

\subsection{Selection of $\phi_a$ and $\phi_b$ for a decomposition penalty}\label{PenlParSel}
We view $\phi_a$ and $\phi_b$ as weights of a tradeoff between preferred and
non-preferred subspaces and assume $\phi_a \cdot \phi_b=constant$. In the current
implementation, we use REML to estimate $\lambda_d$'s for a fixed value of $\phi_a$, and
do a grid search over the $\phi_a$  values to jointly select the tuning parameters which
maximize the information criterion, such as AIC, based on the restricted maximum likelihood.

\section{Simulation} \label{Simulation}
We pursue several simulations to evaluate the properties of the LongPEER method. The
first simulation study (Section~\ref{Simulation1}) compares the performance of the
LongPEER method with the LPFR approach. In the remaining simulation studies, only the
LongPEER method is considered.  The purpose of the second simulation study is to evaluate
the influence of sample size and the contribution of prior information about the
functional structure (as determined by the tuning parameters $\phi_a$ and $\phi_b$ in
\eqref{decomp}) on the LongPEER estimate. In the third simulation study, we evaluate the
coverage probabilities of the confidence bands constructed using the formula presented in
Section~\ref{Precision}. Finally, we evaluate the performance of LongPEER estimate when
information on some features is missing and the results are summarized in
Section~\ref{Simulation4}. In all the simulation studies, the simulated predictor
functions resemble the MRS data. All results summarized in this Section are based on 100
simulated datasets.

For each subject and visit, predictor functions were simulated independently. Predictor
functions were flat with bumps of varied widths at a number of pre-specified locations. White
noise was added to the predictor functions to account for the instrumental measurement
noise. These ``bumpy" regression functions were generated with bumps at some (but, not
all) of the bump locations of the predictor function.  For the simulation in Section
\ref{Simulation1}, the regression function is assumed to be independent of time, whereas
it varies with time in the simulation of Section~\ref{Simulation2}. For both the
predictor and regression functions, 100 equi-spaced sampling points in [0,1] are used.

For the decomposition penalty \eqref{decomp}, the matrix $L_d$ is defined as follows: 1)
select the discretized functions $q_j, j=1, \ldots , J$ spanning the ``preferred"
subspace and 2) compute $L_d=QQ^+$, where $Q=\mbox{col}[q_1, \ldots , q_J]$ and the
vectors $qPj$ are discretized functions, defined to have a single bump corresponding to a
region in the simulated predictor functions; see Figure~\ref{LPFRfig}.  The columns of
$Q$ need not be orthogonal (cf., Figure~\ref{App2}).

Estimation error is summarized in terms of the mean squared error (MSE) of the estimated
regression function defined as $||\gamma - \tilde{\gamma}||^2$, where $\tilde{\gamma}$
denotes the estimate of $\gamma$. Further, MSE was decomposed into the trace of the
variance and squared norm of bias. We also calculated the sum of squares of prediction
error (SSPE) as  $||y - \tilde{y}||^2/N$, where $\tilde{y}$ denotes the estimate of the
true (noiseless) $y$. The estimates based on the proposed methods, including the LongPEER
estimate, were obtained as BLUPs from the mixed model formulation described in Section
\ref{BLUP}.

\subsection{Comparison with LPFR} \label{Simulation1}
As mentioned, LPFR estimates a regression function that does not vary with time.
Therefore, in the first set of simulations we generated outcomes using a time-invariant
regression function (i.e., $\gamma (t,s)=\gamma _0(s)$,  for all $t$). The following
model was used to generate the outcome data for 100 individuals ($i=1, \cdots, 100$),
each at 4 timepoints ($t=0,1,2,3$):
\begin{eqnarray}
y_{it} & = & \beta_0+\int_0^1{W_{it}(s) \gamma_0(s) ds} +b_i +\epsilon_{it},
\qquad i=1,\cdots, 100, \\ \nonumber
\mbox{where, } & & \gamma_0(s) = \sum_{h\in H_{\gamma_0}} {a_{0h}\mbox{ exp}\left[-2500*\Big(\frac{h-s}{100}\Big)^2\right]}.
\label{eqn:gamma0}
\end{eqnarray}
The bumpy predictor functions were generated from the following equation
\begin{eqnarray}
w_{it}(s) & = & \sum_{h\in H_1}{(\xi_{1h}+c_{1h})\mbox{exp}\left[-2500*\left(\frac{s-h}{100}\right)^2\right]}\\
&& + \sum_{h\in H_2}{(\xi_{2h}+c_{2h})\mbox{exp}\left[-1000*\left(\frac{s-h}{100}\right)^2\right]}\nonumber \\
           &&     + \ (\xi_{31} +0.9)\mbox{exp}\left[-250*\left(\frac{s-50}{100}\right)^2\right], \nonumber
\end{eqnarray}
where $c_{1h}$, $c_{2h}$ and $a_{0h}$ are defined in Table \ref{LPFRtab1}. $\{\xi_{1h}, h
\in H_1\}$, $\{\xi_{2h}, h \in H_2\}$, and $\xi_{31}$ were drawn independently from
Uniform(0, 0.1). Also, $\beta_0 = 0.06$,  $\epsilon_{it} \sim N[0, (0.02)^2]$ and $b_i
\sim N[0, (0.05)^2]$.

\begin{table}[h]
\caption{Values of $c_{1h}$, $c_{2h}$, $a_{0h}$ and $a_{1h}$ for generating predictor and
regression function in simulation studies in Sections~\ref{Simulation1},
\ref{Simulation2}, \ref{Simulation3} and \ref{Simulation4}.} \label{LPFRtab1}
\centering
\begin{tabular}
{cc|cc|cc|cc}
\hline
\multicolumn{2}{c|}{$h \in H_1$} & \multicolumn{2}{c|}{$h \in H_2$}
& \multicolumn{2}{c|}{$h \in H_{\gamma_0}$} & \multicolumn{2}{c}{$h \in H_{\gamma_1}$} \\
\hline
 $ h$ & $c_{1h}$ &  $ h$ & $c_{2h}$  &  $ h$ & $a_{0h}$ &  $ h$ & $a_{1h}$ \\
\hline
15 & 0.10 & 30 & 0.60 & 15 & 0.20 & 30 & 0.06 \\
5 & 0.10 & 70 & 0.50 & 50 & -0.15 & 70 & -0.06 \\
& &  80 & 0.50 & 80 & 0.15 &&\\
&& 90 & 0.40 &&&&\\
\hline
\end{tabular}
\bigskip
\end{table}

\begin{figure}
\centerline{%
\includegraphics [angle=0,width=135mm, height=50mm]{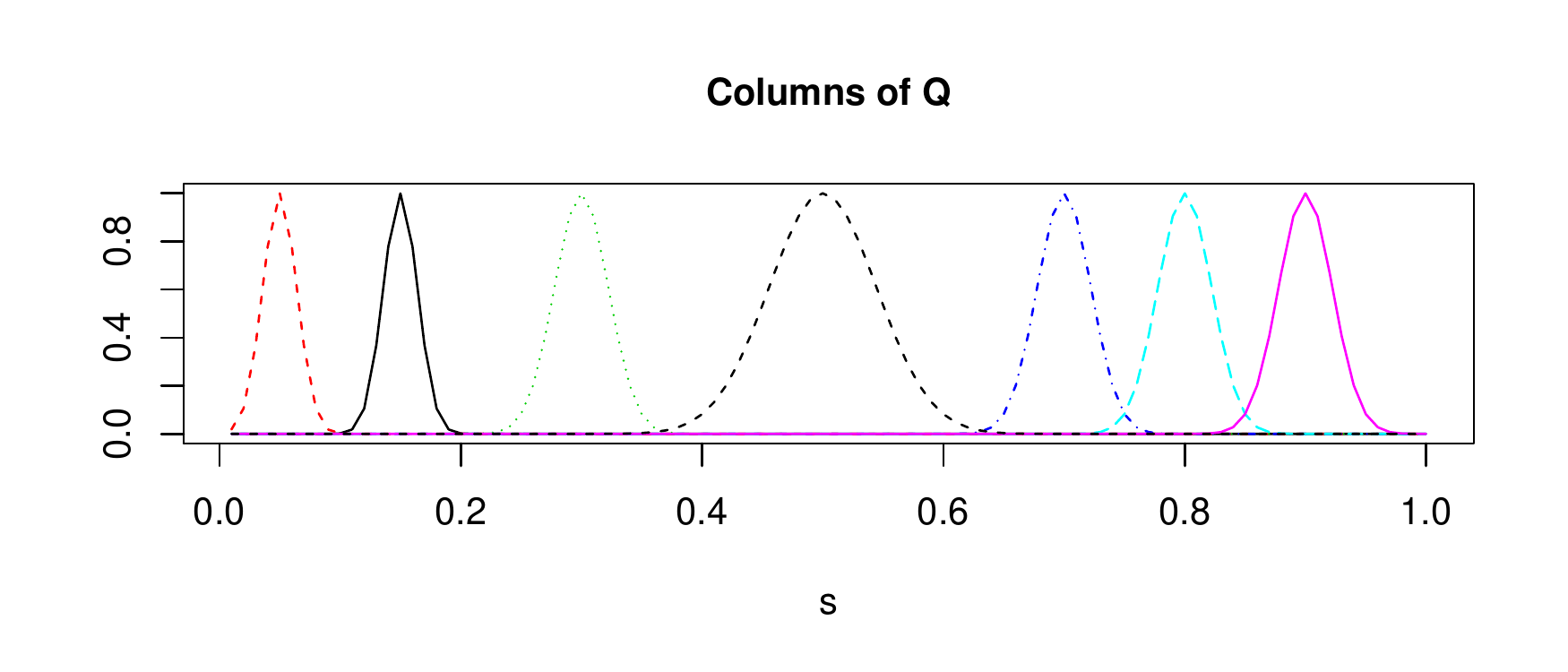}}
\centerline{%
\includegraphics [angle=0,width=135mm, height=70mm]{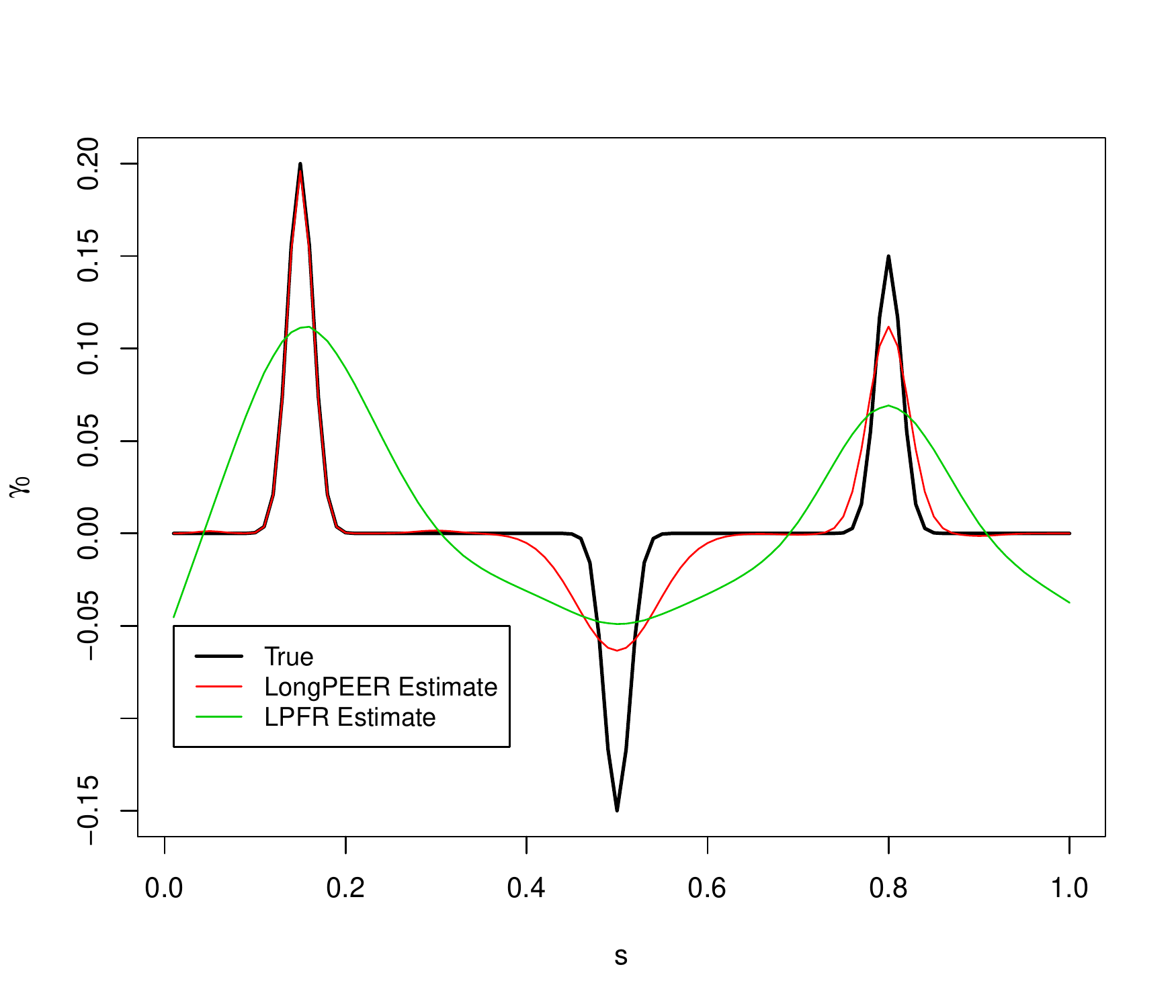}}
\caption{Average estimates of $\gamma$ for the simulation in Section \ref{Simulation1} with $\phi_a=10$ and $\phi_b=1$.
{\it Top panel}: columns of $Q$ used in the decomposition penalty. {\it Bottom panels}: the true $\gamma$ and
the average of estimates from 100 simulations.}
\label{LPFRfig}
\end{figure}

We applied both LPFR (using \verb|lpfr()| available in the \verb|refund| package in
\verb|R| \citep{refund}) and the LongPEER method to the simulated data. To obtain the
LPFR estimate, the dimension of both principal components for predictor function and
truncated power series spline basis for the regression function were set to 60. The
columns of $Q$ used to define $L_Q$, for the LongPEER estimate are plotted in the top
panel of Figure~ \ref{LPFRfig}. We used $\phi_a/\phi_b=10$, a choice motivated by our
findings in Sections~\ref{Simulation2} and \ref{Simulation4}.

Table~\ref{LPFRtab} displays the MSE and prediction error obtained for LongPEER and LPFR
estimates. The SSPE was similar for both methods (1.1566 and 1.1535), however, the
LongPEER estimate has smaller MSE. Both the bias and variance are higher for the LPFR
estimate and consequently it has the greater MSE. Figure~\ref{LPFRfig} displays the
estimates of the regression function.
It should be emphasized that any comparison of these methods is not entirely fair since
LongPEER is designed to exploit presumed structural information while LPFR is not.
We note also that the ability to exploit such information may be limited and so in this
simulation we used imprecise information about the shapes of features; see top panel in
Figure~\ref{LPFRfig}. Not surprisingly, performance is best for the feature at $s=0.15$
where information about the shape was relatively precise. See also
Section~\ref{Simulation4}.

\begin{table}
\caption{Estimation and prediction errors for LPFR and LongPEER estimates based on 100 simulated datasets.
The sample size is $N=100$ and the number of longitudinal observations is $n_i=4$.} \label{LPFRtab}
\centering
\begin{tabular}
{l|c|c}
\hline
& LongPEER & LPFR \\
\hline
\hline
$\mbox{MSE}(\gamma_0)$&0.0323&0.2244\\
$\verb| |\mbox{Trace of Variance}(\gamma_0)$&0.0028&0.0490\\
$\verb| |||\mbox{Bias}(\gamma_0)||^2$&0.0295&0.1754\\
$\mbox{SSPE of } Y $&1.1566&1.1535\\
\hline
\end{tabular}
\end{table}

\subsection{Simulation with a time varying regression function}\label{Simulation2}
\begin{figure}
\centerline{%
\includegraphics [angle=0,width=135mm, height=70mm]{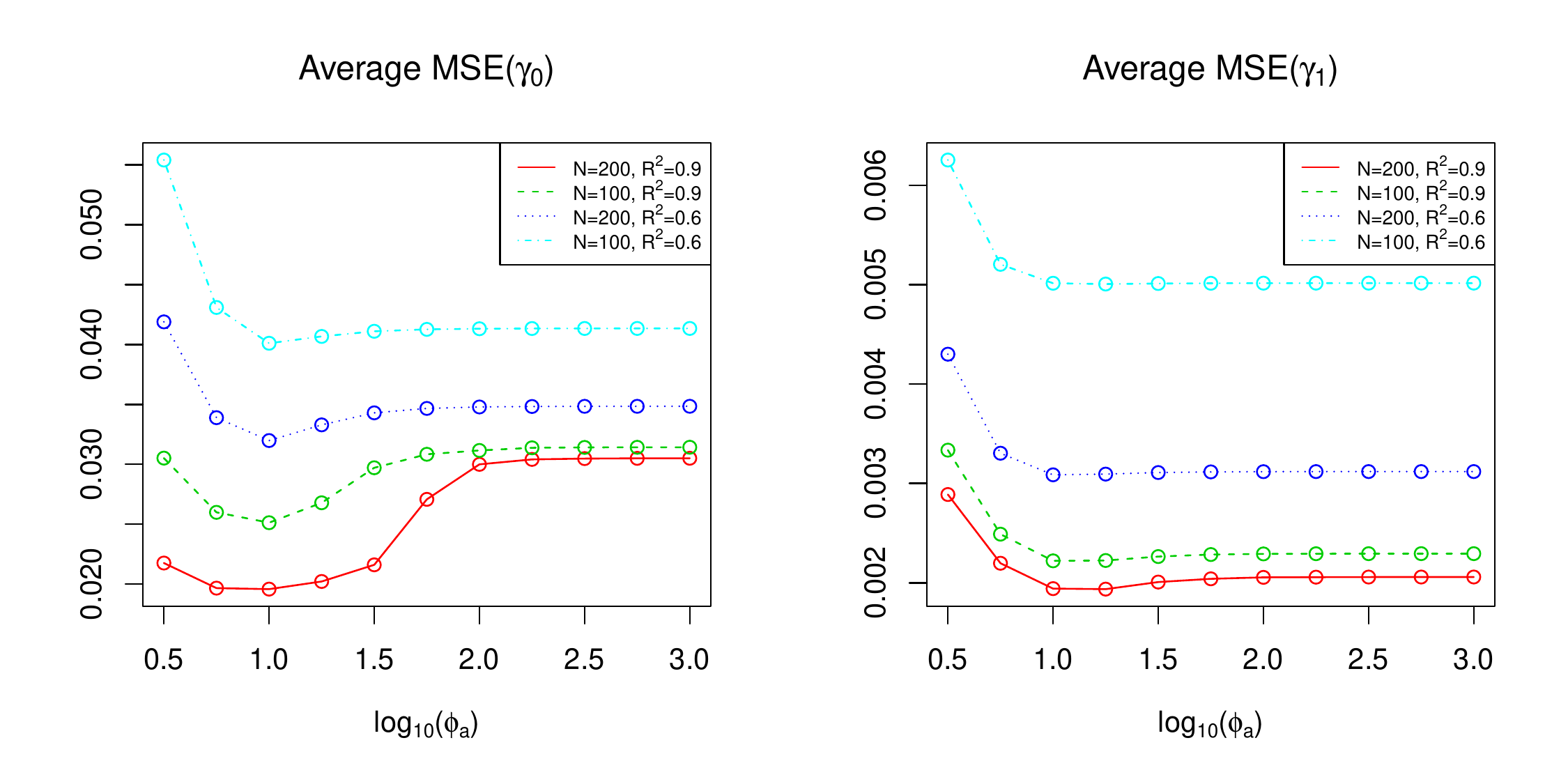}}
\centerline{%
\includegraphics [angle=0,width=135mm, height=60mm]{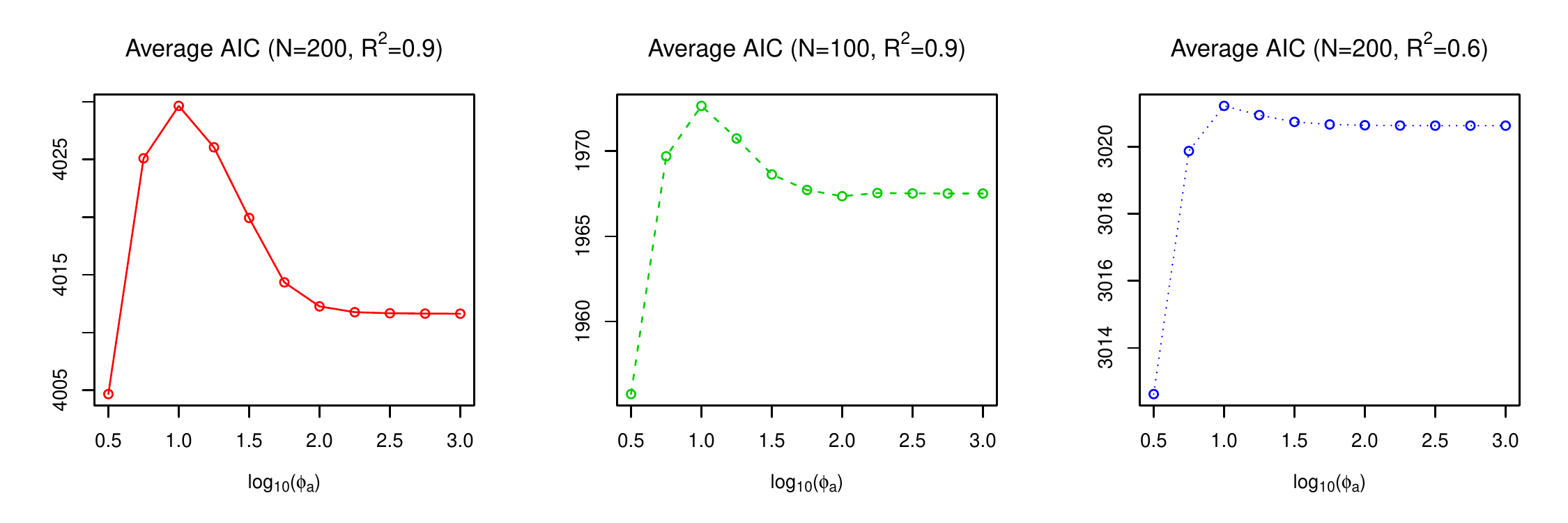}}
\centerline{%
\includegraphics [angle=0,width=135mm, height=60mm]{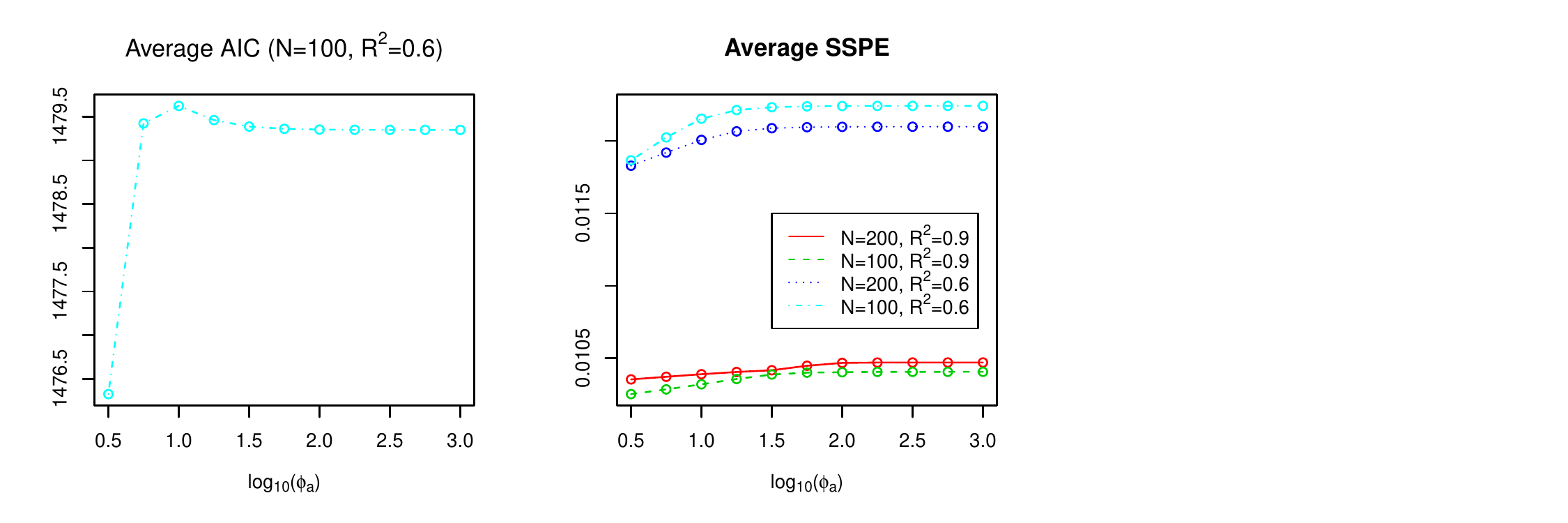}}
\caption{Average AIC, SSPE and MSE for simulations in Section~\ref{Simulation2} over 100 simulations.
At $\phi_a=10$, average AIC were maximized  and MSE($\gamma_0$) and MSE($\gamma_1$) were minimized.
In general, average AIC increased with the increase in sample size and $R^2$ whereas SSPE,
MSE($\gamma_0$) and MSE($\gamma_1$) decreased. }
\label{Sim2.1fig}
\end{figure}

Here the regression function varies parametrically with time. Lacking other functional
regression methods that estimate a time-varying regression function, we only evaluated
the performance of LongPEER. The primary goal was to assess the effects of sample size,
fraction of variance  explained by the model, and the relative contribution of external
information (as determined by $\phi_a$ and $\phi_b$ in equation~\ref{decomp}) on
estimate.

Without loss of generality, we set $\phi_b=1$ and vary $\phi_a$ on an exponential scale.
Larger values of $\phi_a$ indicate greater emphasis of prior information on the
estimation process. The model considered here is similar to that described in
Section~\ref{Simulation1} with the exception that $\gamma (t,s)=\gamma _0(s) + t \;
\gamma _1(s)$. The function $\gamma_0(s)$ is defined in equation (\ref{eqn:gamma0}) and
$\gamma_1(s)$ is of the form
\begin{equation*}
\gamma _1(s)=\sum_{h\in H_{\gamma_1}} {a_{1h}\; \mbox{exp}\left[-2500*\left(\frac{h-s}{100}\right)^2\right]}
\end{equation*}
where the value of $h$ and $a_{1h}$ are listed in Table \ref{LPFRtab1} and  $\beta_0=
0.06$. Realizations of functional predictors were generated as described in
section~\ref{Simulation1}. For each simulation, an appropriate $\sigma_{\epsilon}^2$ was
chosen to ensure that the squared multiple correlation coefficient $R^2=s_y^2/[s_y^2+
\sigma_{\epsilon}^2]$ is $0.6$ and $0.9$. Here,
$s_y^2=\frac{1}{4}\sum_{t=0}^3{\frac{1}{N-1}\sum_{i=1}^N{(y_{it}-\bar{y}_{.t})^2}}$
denotes the average sample variance in the set $\{y_{it}-\epsilon_{it}:i=1, \cdots, N;
t=0, \cdots, 3\}$ with $\bar{y}_{.t}=\frac{1}{N}\sum_{i=1}^N{y_{it}}$.

\begin{figure}
\centerline{%
\includegraphics [angle=0,width=135mm, height=70mm]{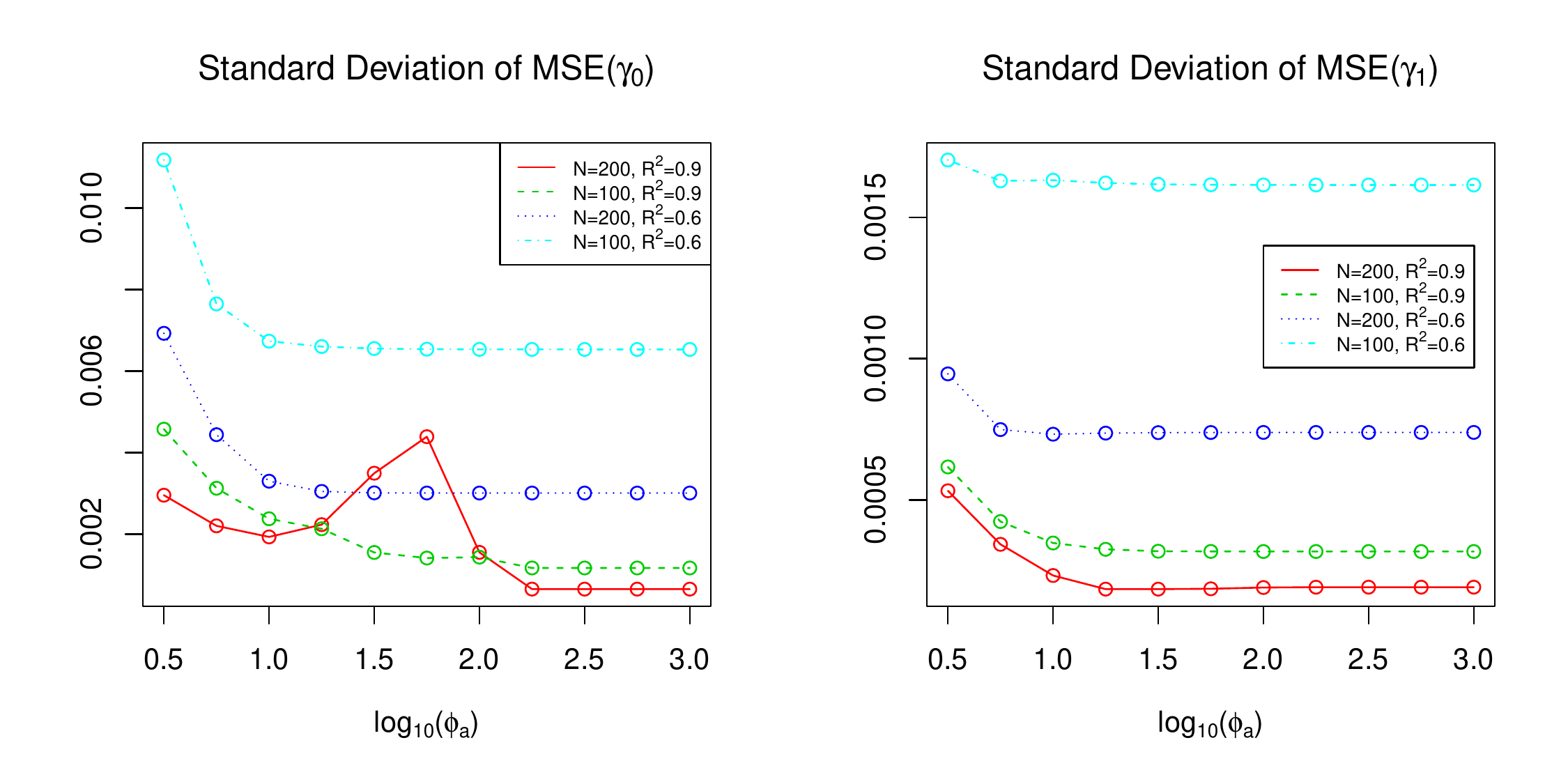}}
\caption{Standard deviation of MSE for simulations in Section~\ref{Simulation2} over 100 simulations.
Standard deviation of MSE($\gamma_0$) and MSE($\gamma_1$) generally decrease with increasing sample size and $R^2$.
MSE($\gamma_0$) and MSE($\gamma_1$) both decrease up to $10^1.25$ and then plateau, except for MSE($\gamma_0$)
in the scenario with $N=200$ and $R^2=0.9$.}
\label{Sim2.1fig.MSE}
\end{figure}

\begin{figure}
\centerline{%
\includegraphics [angle=0,width=135mm, height=50mm]{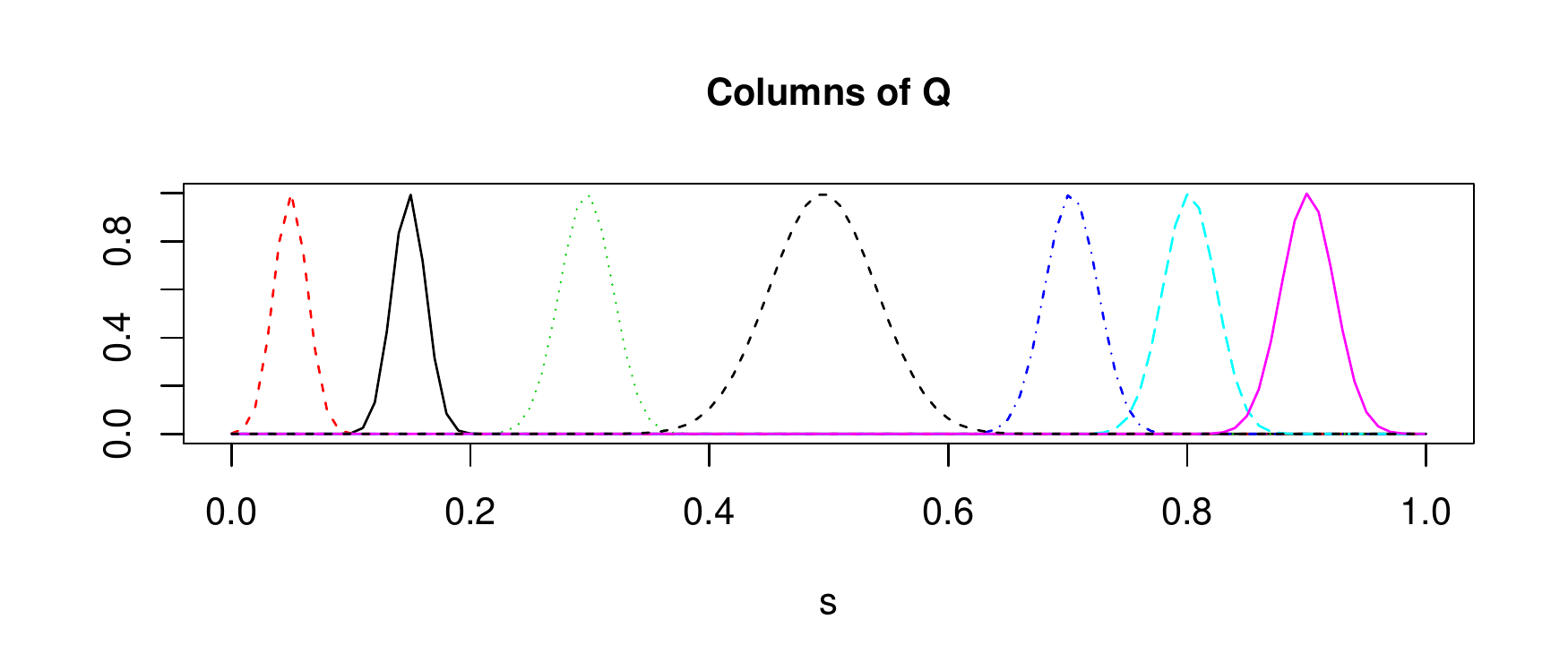}}
\centerline{%
\includegraphics [angle=0,width=135mm, height=140mm]{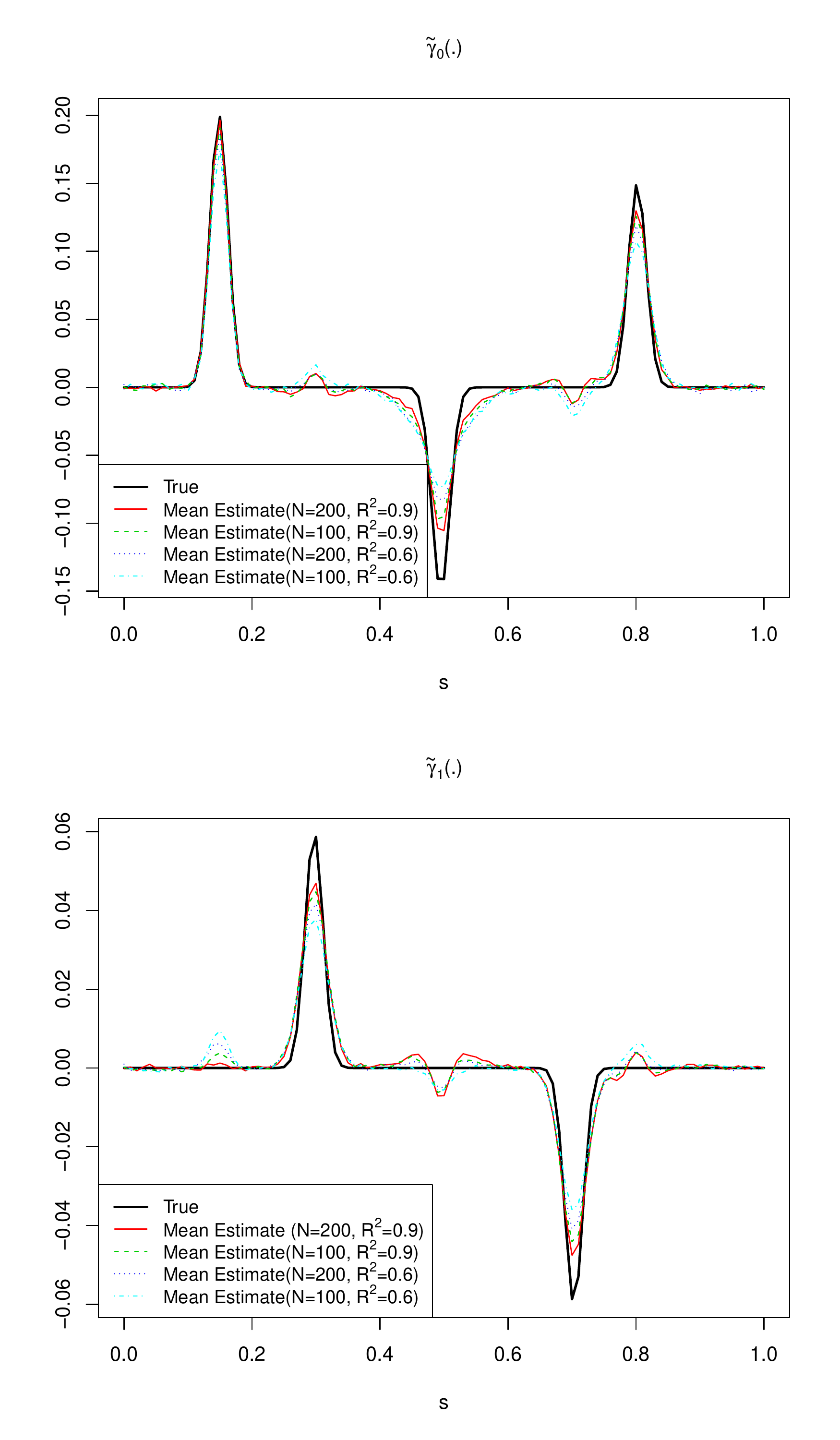}}
\caption{Average estimates of the components of regression functions for simulations described in
Section~\ref{Simulation2} with $\phi_a=10$ and $\phi_b=1$. {\it Top panel}: columns of $Q$ used in defining
a {\it decomposition penalty}. {\it Middle and bottom panels}: the average estimates of $\gamma_0$ and $\gamma_1$;
these improve as $N$ and/or $R^2$ increase.}
\label{Sim2.2fig}
\end{figure}

We have repeated the simulation for four scenarios: (i) $N=100$, $R^2=0.6$; (ii) $N=100$,
$R^2=0.9$; (iii) $N=200$, $R^2=0.6$; and (iv) $N=200$, $R^2=0.9$. Estimate of $\gamma_0$
and $\gamma_1$ were obtained using a decomposition penalty. The columns of $Q$ used to
define $L_Q$  are plotted in the top panel of Figure~\ref{Sim2.2fig}. Results for AIC,
MSE and SSPE are displayed graphically in Figure~\ref{Sim2.1fig}. The standard deviation
of MSE were plotted in Figure~\ref{Sim2.1fig.MSE}. As the sample size and $R^2$
increased, both the MSE$(\gamma_0)$ and MSE $(\gamma_1)$ were decreased, providing
empirical evidence that the LongPEER estimates were consistent. In all four scenarios,
MSE($\gamma_0$) was minimized at $\phi_a=10$, it increased with $\phi_a$ up to
$\phi_a=100$, and plateaued after that.  On the other hand, a decrease in MSE($\gamma_1$)
is observed as $\phi_a$ increased up to 10 and it plateaued thereafter. That is, an
increase in $\phi_a$ up to 10 resulted in improvement in estimation of both $\gamma_0$
and $\gamma_1$. However, $\phi_a$ beyond 10 resulted in deterioration in performance of
estimation for $\gamma_0$; estimation performance for $\gamma_1$ remained almost
unchanged. To understand this result, we need to compare the plots of columns for $Q$
matrix used in defining $L_Q$ with true $\gamma_0$ and $\gamma_1$ in
Figure~\ref{Sim2.2fig}: $\gamma_0$ has peaks at $s=0.2$, $0.5$ and $0.8$. and $Q$
contains functions (colums) representing peaks at these locations. However, the shape of
the peak at $s=0.5$ is different from that in $\gamma_0$. Due to this difference in
shape, as $\phi_a$ increased  from 10 to 100, the feature at $s=0.5$ in
$\tilde{\gamma}_0$ became smaller leading to gradual increase in MSE($\gamma_0$). On the
other hand, $\gamma_1$ has two features while $Q$ contains functions of very similar
shape. Consequently, MSE($\gamma_1$) stabilizes after $\phi_a=10$.
Finally, note that the value of $\phi_a$ that maximized AIC also minimized
MSE$(\gamma_0)$ and MSE$(\gamma_1)$. This suggests that AIC can be used to guide the
choice of $\phi_a$ while setting $\phi_b$ at 1. In general, the choice of $\phi_a$ may be
take as that which maximizes AIC.

The average LongPEER estimate of $\gamma_0$ and $\gamma_1$ using a decomposition penalty
are displayed in Figure~\ref{Sim2.2fig} with $\phi_a=10$ and $\phi_b=1$. For smaller
sample sizes and $R^2$, the LongPEER estimate may: (a) oversmooth (i.e., negatively
bias) the estimated regression function at locations of a true feature, and (b) be
positively biased in locations corresponding to features in $\mathcal{Q}$
but where the true $\gamma$ is zero. However, by increasing the sample size to 200 and/or
$R^2$ to 0.9, we observe that the average LongPEER estimate $\gamma_0(\cdot)$ and
$\gamma_1(\cdot)$ approach the true functions.

\subsection{Coverage probability} \label{Simulation3}

\begin{figure}
\centerline{%
\includegraphics [angle=0,width=135mm, height=40mm]{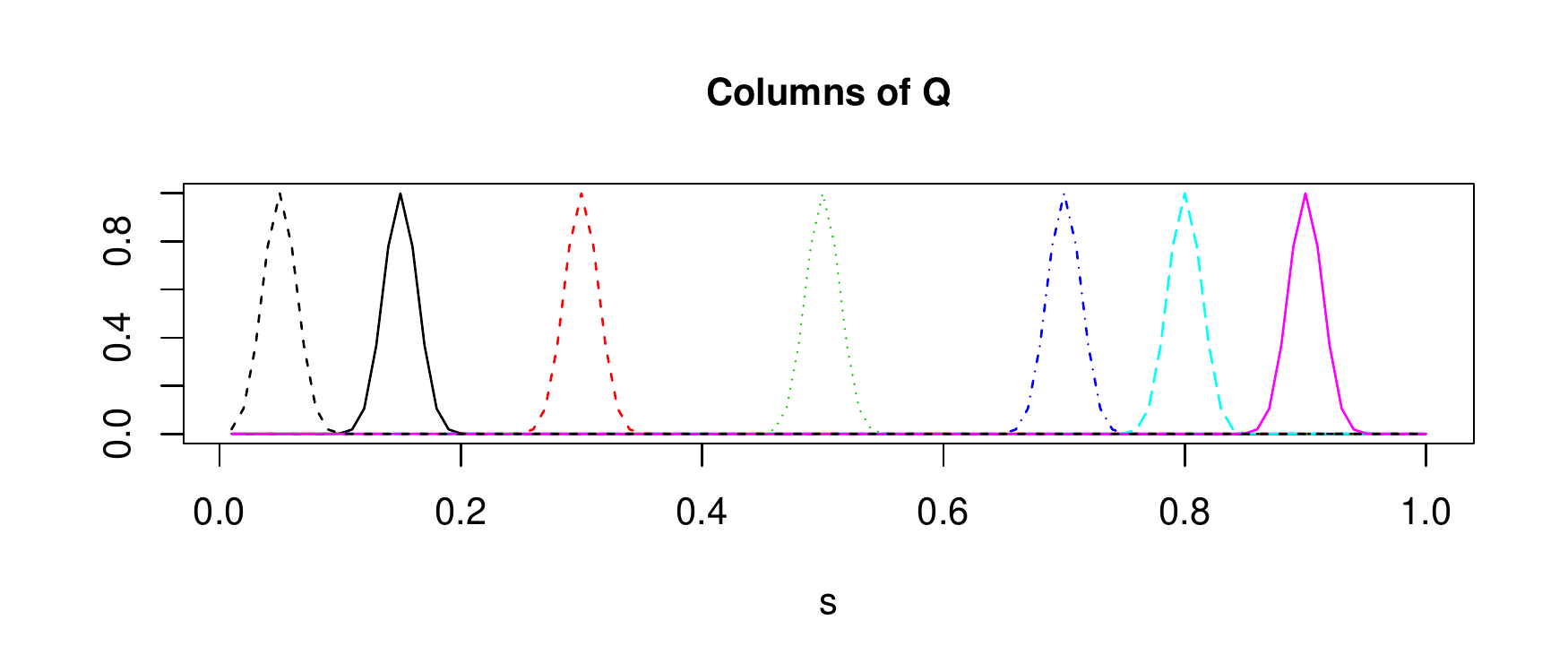}}
\centerline{%
\includegraphics [angle=0,width=135mm, height=120mm]{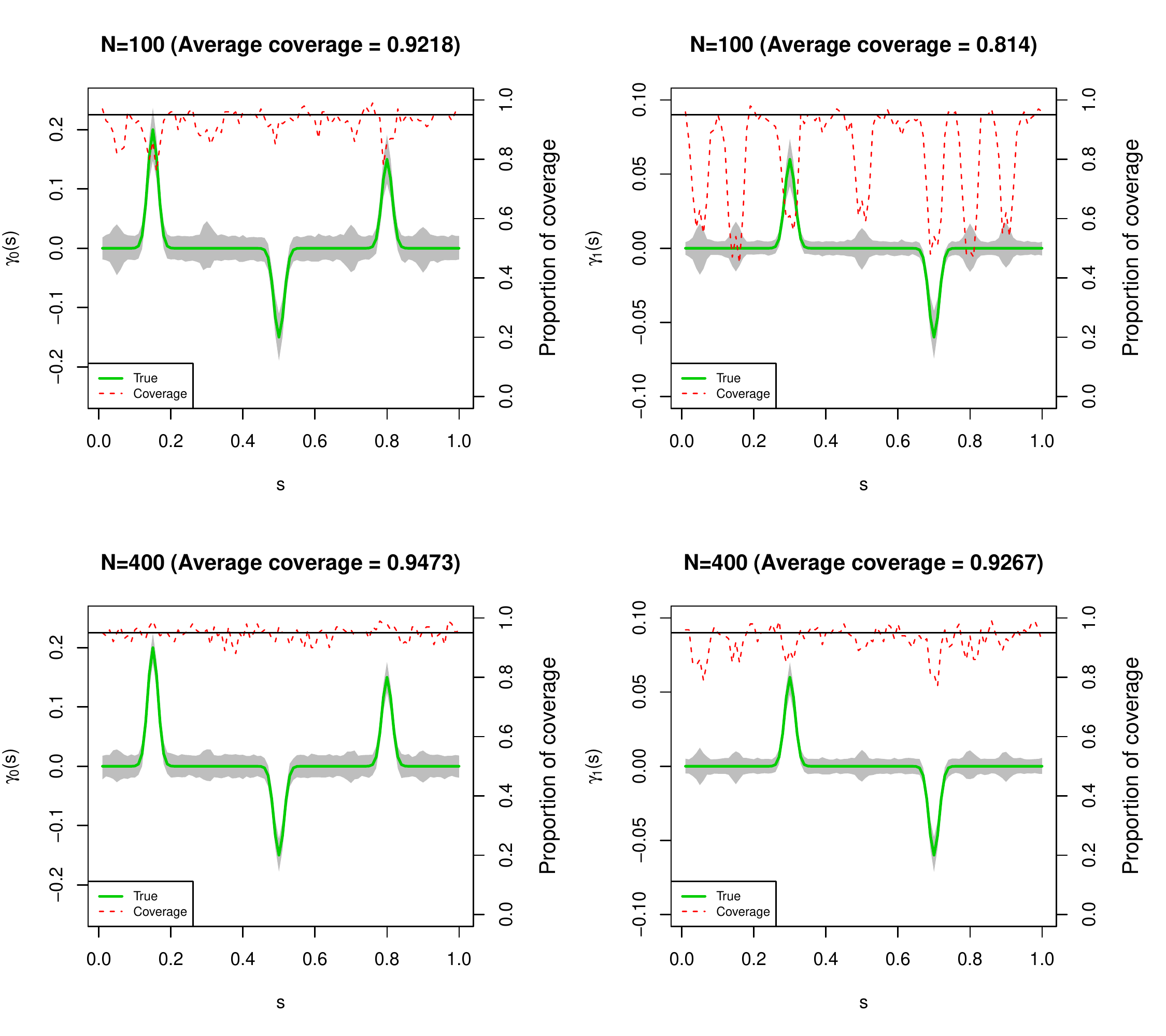}}
\caption{Coverage probabilities of LongPEER estimates in 100 simulations with $\phi_a=10$ and $\phi_b=1$
discussed in section \ref{Simulation3}. {\it Top panel}: the columns of $Q$ used in the {\it decomposition penalty}.
{\it Middle and bottom panels}: pointwise 95\% confidence band (shaded region) and coverage
proportions (the dotted line) based on $N=100$, and $N=400$ subjects, respectively. The left column displays
 the cross-sectional function $\gamma_0(\cdot)$ and the right column the longitudinal function $\gamma_1(\cdot)$.
 The horizontal line in each plot marks the nominal coverage of 95\%.}
\label{Sim2.3fig}
\end{figure}
In this section, we used the simulation setup described in Section \ref{Simulation2} with
$R^2=0.9$. The columns of $Q$ matrix used in defining the decomposition penalty
\eqref{decomp} are displayed in the top panel of Figure \ref{Sim2.3fig}. The middle and
bottom panel shows the confidence bands and the coverage probabilities obtained using
$\phi_a=10$. The 95\% confidence bands are constructed as $Estimate$  $\pm 1.96\times$
$(Standard$ $Error)$. When the sample size $N$ increased, there was a notable improvement
in coverage of both $\gamma_0(\cdot)$ and $\gamma_1(\cdot)$. For $N=100$,  the coverage
of $\gamma_1(\cdot)$ by the confidence bands was only around 81\%. This confidence band
under-coverage of $\gamma_1(\cdot)$ is caused by the comparatively larger bias in the
estimation of $\gamma_1(\cdot)$ with $N=100$ (see Section~\ref{Simulation2} and
Figure~\ref{Sim2.2fig}). The observed coverage increases with $N$: for $N=400$, the
coverage is very close to 95\%.  We also explored the influence of $\phi_a$ on the
confidence band and coverage probability (not shown here). The higher values of $\phi_a$
led to the confidence band shrinkage and this in turn resulted in under-coverage of both
$\gamma_0(\cdot)$ and $\gamma_1(\cdot)$.

\subsection{Estimation in the presence of incomplete information} \label{Simulation4}

\begin{figure}
\centerline{%
\includegraphics [angle=0,width=135mm]{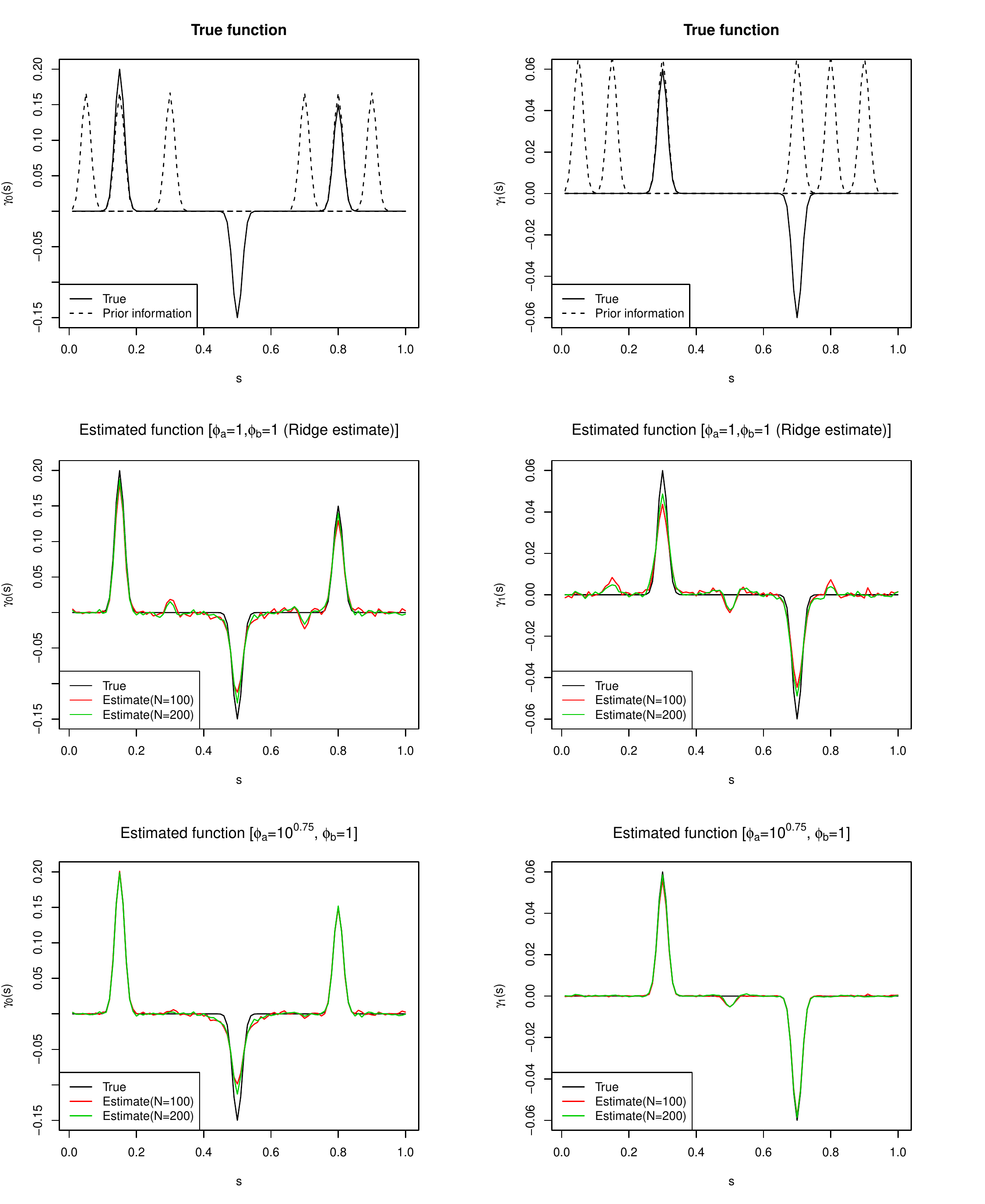}}
\caption{{\it Top panel}: true regression functions (solid lines) $\gamma_0(\cdot)$ (left) and $\gamma_1(\cdot)$ (right)
and 6 vectors spanning $P_Q$ (dashed lines). {\it Middle panel}: Average ridge-penalty estimate
from 100 simulations. {\it Bottom panel}: Average LongPEER estimate from 100 simulations with $P_Q$
defined by the 6 vectors displayed in the top panel and $\phi_a=10^{0.75}$.} \label{Sim2.4fig}
\end{figure}

Since the LongPEER estimate uses external information in the estimation process, it is of
interest to evaluate its estimation performance when only partial information is
available. In this section, we use a simulation scenario similar to that in
Section~\ref{Simulation3}, but now the penalty is defined without regard for information
about the peak at $s=0.5$.  As displayed in Figure \ref{Sim2.4fig}, the LongPEER
estimates of $\gamma_0(s)$ has appropriate structure at $s=0.5$, on average. Indeed as
with an ordinary ridge penalty, this structure is inherited from the empirical
eigenvectors of $W(\cdot)$. This highlights the advantage of an estimate obtained from
the jointly-determined eigenvectors of $W(\cdot)$ and $L$ (see Appendix \ref{GSVD}); the
estimate depends on the relative contributions $W$ and $L$, controlled by the ratio of
$\phi_a$ to $\phi_b$.

The relative increase in the contribution of external information in the estimation
process resulted in shrinkage towards zero at $s=0.5$.  The estimates displayed in
Figure~\ref{Sim2.4fig} result from $\phi_b=\phi_a=1$ (i.e., a ridge penalty) in the
middle panel, and $\phi_b=1$, $\phi_a=10^{0.75}$ in the bottom panel. For values of
$\phi_a$ larger than $10^{0.75}$, minimal changes in the estimates are observed.

\subsection{MRS study application} \label{Application}

\begin{table}
\caption{Comparison of AIC for selection of scalar covariates, $\phi_a$ ($\phi_b=1$) and time structure in $\gamma(t, \cdot)$ in Section~\ref{Application}}
\label{AICmodel}
\begin{tabular}{l|l|l|r|r}
\hline
& Scalar covariates & Time structure in $\gamma(t, \cdot)$ & $\phi_a$  & AIC\\
\hline
Model 1& $t$ & $\gamma_0(t, \cdot) + t \gamma_1(t, \cdot)$ & 10 & $-395.2335$\\
Model 2& Age, $t$ & $\gamma_0(t, \cdot) + t \gamma_1(t, \cdot)$ & 10 & $-405.2796$\\
Model 3& Gender, $t$ & $\gamma_0(t, \cdot) + t \gamma_1(t, \cdot)$ & 10 & $-395.9040$\\
Model 4& Race, $t$ & $\gamma_0(t, \cdot) + t \gamma_1(t, \cdot)$ & 10 & $-398.5607$\\
Model 5& $t$, $t^2$& $\gamma_0(t, \cdot) + t \gamma_1(t, \cdot)+ t^2 \gamma_2(t, \cdot)$ & 10 & $-394.5752$\\
Model 6& $t$ & $\gamma_0(t, \cdot) + t \gamma_1(t, \cdot)$ & 100 & $-395.367$0\\
\hline
\end{tabular}
\end{table}

\begin{figure}
\centerline{%
\includegraphics [angle=0,width=140mm, height=60mm]{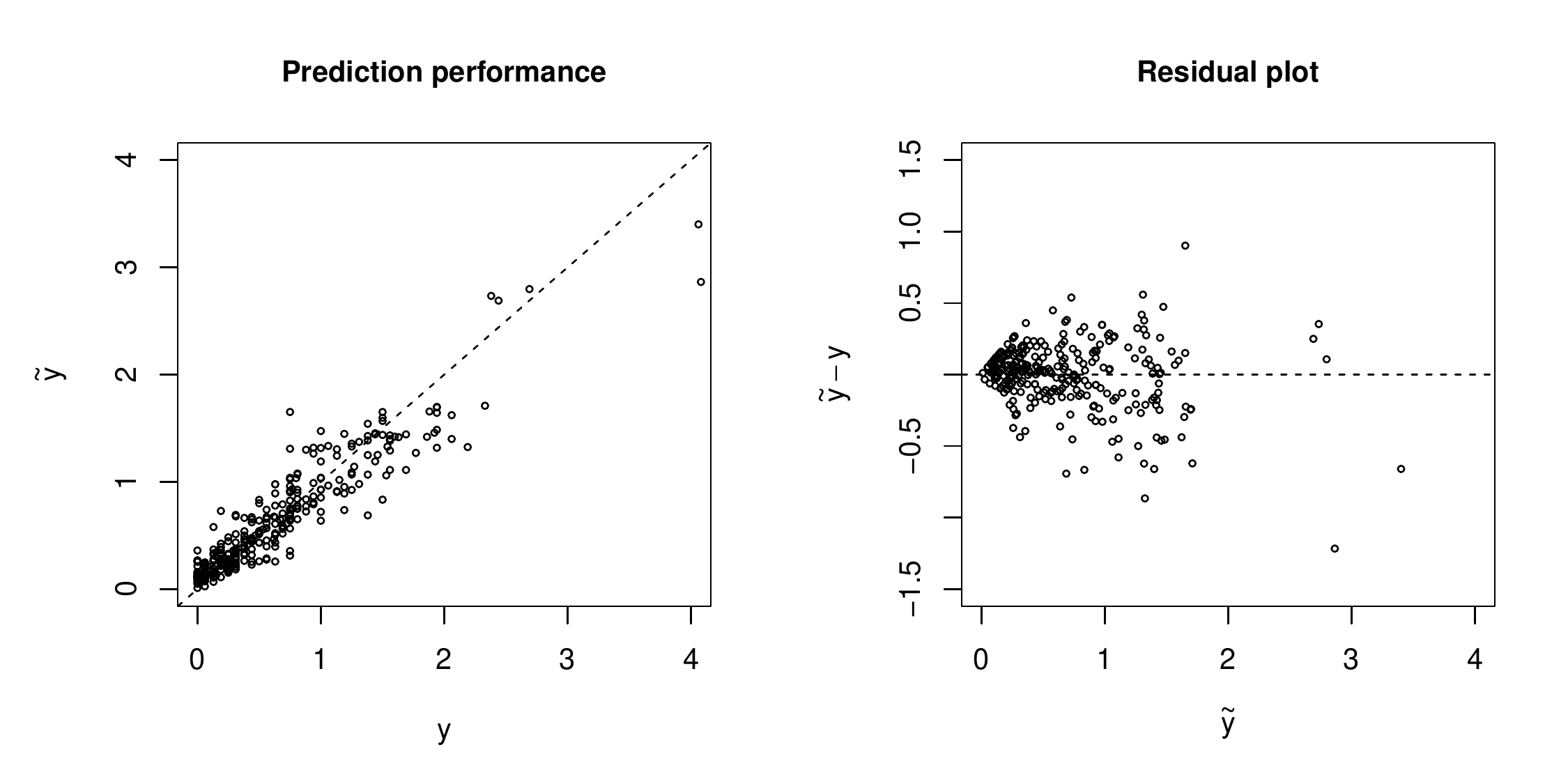}}
\caption{Prediction performance of Model in equation~\eqref{RealModel}.
Left panel: observed GDS score ($y$) and predicted value ($\tilde{y}$). Right panel: observed
$\tilde{y}$ and residuals $(y-\tilde{y})$.}
\label{App2diagnostic}
\end{figure}

\begin{figure}
\centerline{%
\includegraphics [angle=0,width=80mm, height=80mm]{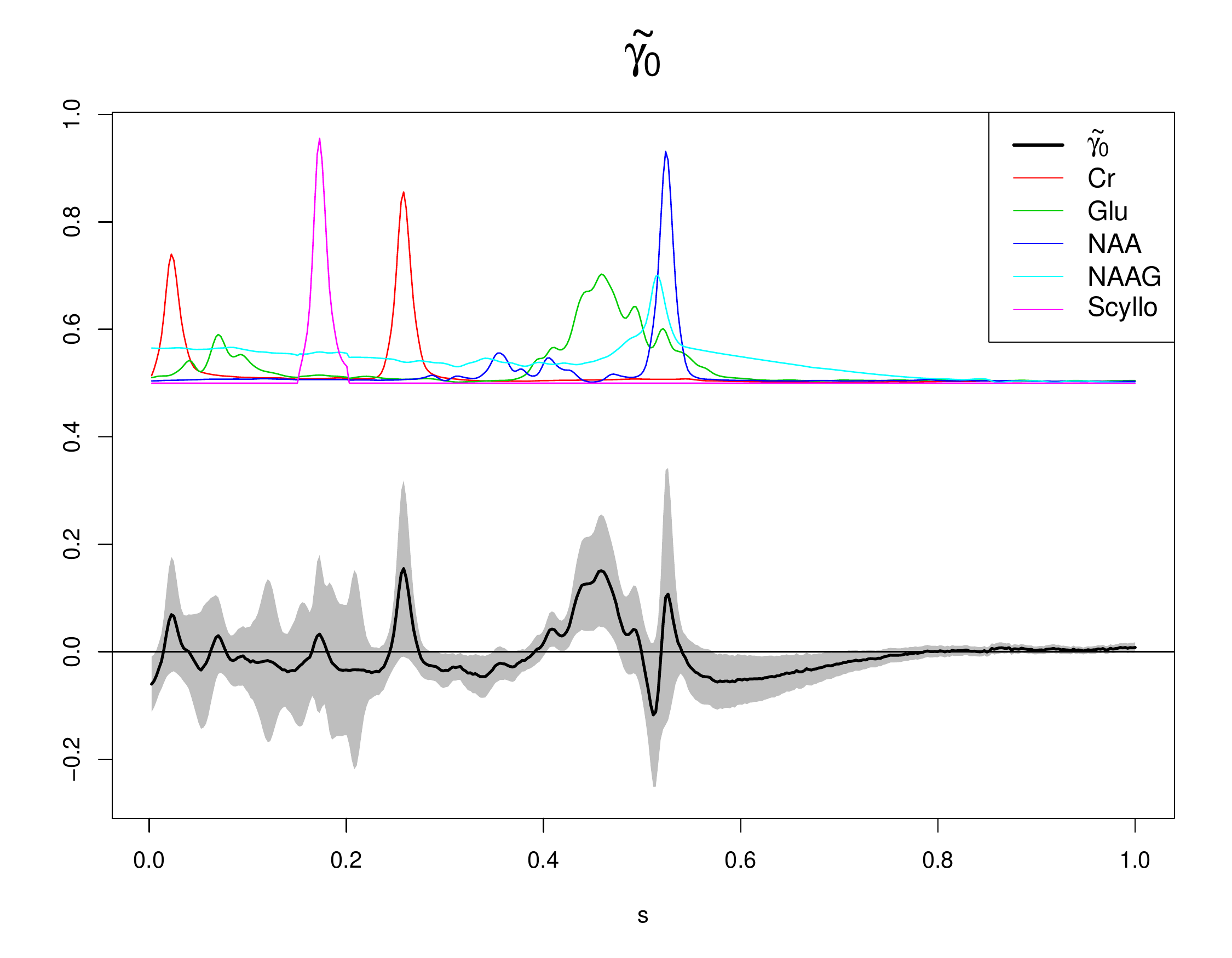}}
\centerline{%
\includegraphics [angle=0,width=80mm, height=80mm]{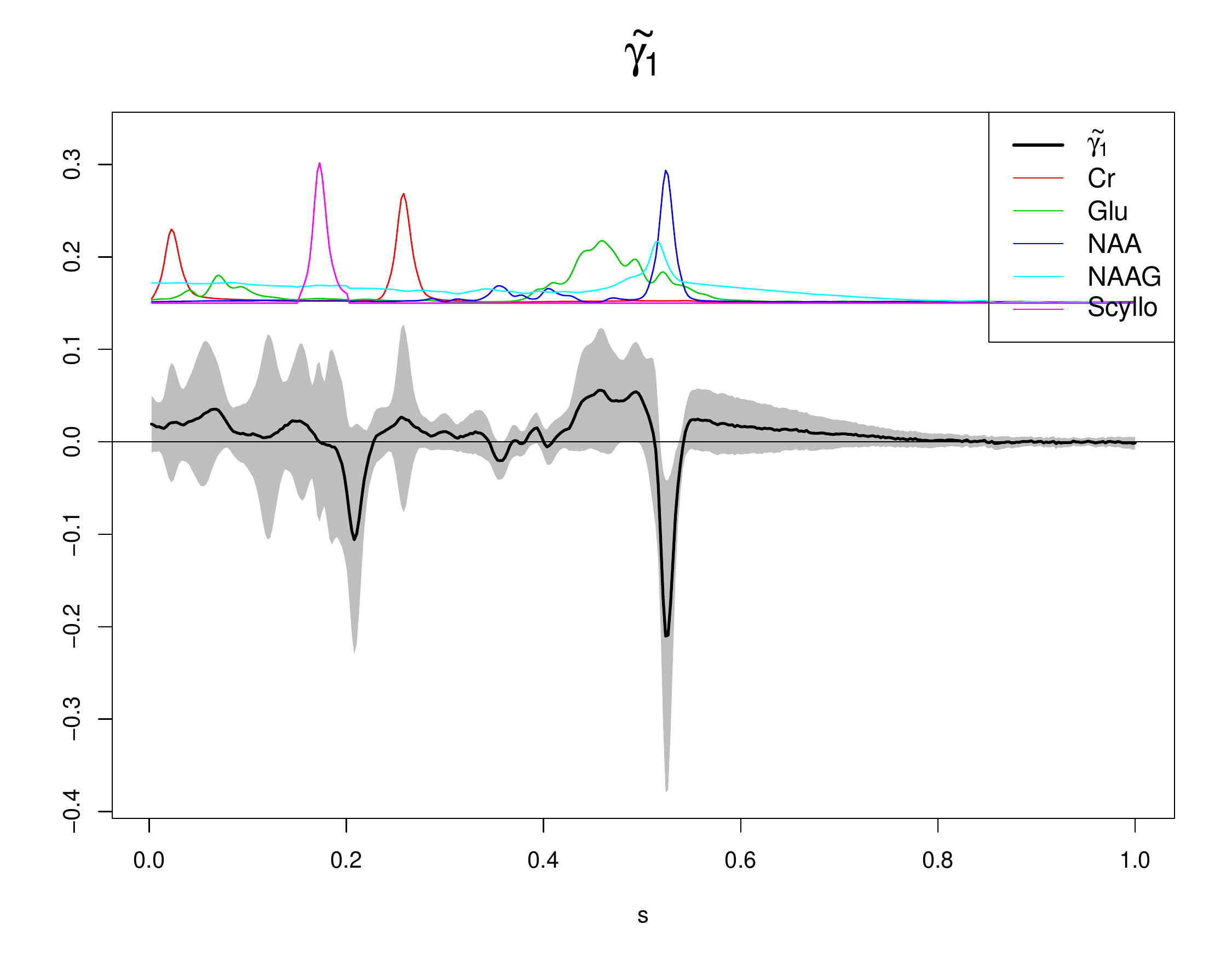}}
\caption{Estimates of the regression function (with 95 \% pointwise confidence band) as
described in Section \ref{Application}.  Shaded region in both the plots represent pointwise
confidence bands. Top panel: estimate of $\gamma_0(\cdot)$. Bottom panel: estimate of $\gamma_1(\cdot)$.
Selected (scaled) pure metabolite spectra are also shown on both plots.
Estimation used a decomposition penalty with $\phi_a= 100$, $\phi_b=1$.}
\label{App2}
\end{figure}

We applied LongPEER to investigate potential associations of metabolite spectra, obtained
from basal ganglia, and the global deficit score (GDS) in a longitudinal study of late
stage HIV patients. Of particular interest is how such an association evolves over
time. The study description is available elsewhere \citep{Harezlak2011}. We treat global
deficit score (GDS)  as our scalar continuous response variable  and MR spectrum (sampled
at $K=399$ distinct frequencies) as  functional predictor. GDS is often used as a
continuous measure of neurocognitive impairment \citep[e.g.,][]{Carey2004} and a large
GDS score indicates a high degree of impairment. The MRS spectra are comprised  of pure
metabolite spectra, instrument noise and a background profile. We collected a total of
$n_{\bullet}=306$ observations from $N=114$ subjects. The longitudinal observations for
each subject were within 3 years from baseline. The number of observations per subject
ranged from 1 to 5 with a median equal to 3. Spectral information of 9 pure metabolites
was used as prior information for the LongPEER estimation.  The pure metabolite spectra
are: Creatine (Cr), Glutamate (Glu), Glucose (Glc), Glycerophosphocholine (GPC),
\textit{myo}-Inositol (Ins), N-Acetylaspartate (NAA), N-Acetylaspartylglutamate (NAAG),
\textit{scyllo}-Inositol (Scyllo) and Taurine (Tau). These spectra are displayed in
Figure \ref{fig:App1}. The decomposition penalty, $L_Q$, defined as in
equation~\eqref{decomp} where $Q=[q_1, \cdots, q_9]$, is a matrix of dimension
$9\times399$.

Information available on demographic factors includes: {\it age} at baseline, {\it
gender} and {\it race}. We relied on AIC to  choose (a) scalar covariates in the model,
(b) $\phi_a$ (while setting $\phi_b= 1$) for defining decomposition based penalty $L_Q$
and (c) the time structure of $\gamma(t, \cdot)$. Based on the AIC (see
Table~\ref{AICmodel}),  Models 1, 3, 5 and 6 are almost identical and appear to be better than the remaining models. In
these models, $\phi_a$ was selected to be either $10$ or $100$  and gender is the only scalar covariate.
Models 1 and 5 were different with respect to time structure in $\gamma (t,s)$.  Although including
$\gamma_2(t,\cdot)$ led to a marginal increase in AIC ($-394.58$ vs $-395.23$), we
did not observe any significantly non-zero region of $\gamma_2(t, \cdot)$, based on
pointwise 95\% confidence intervals in Model 5.
Models 1 and 6 were different in terms of the $\phi_a$. Use of smaller $\phi_a$ led to slight increase in AIC ($-395.23$ vs $-395.37$). However, the interpretability of the estimates for $\gamma_0(\cdot)$ and $\gamma_1(\cdot)$ became harder because of their increased wiggliness leading to the choice of Model 6.

Hence, we fit Model 1 (with $\phi_a= 100$,$\phi_b= 1$) as follows:
\begin{equation}
y_{it}= \beta_0 + \beta_1 \; t +\int_{\Omega}{W_{it}(s) \gamma (t,s)ds} + b_i +\epsilon_{it}, \label{RealModel}
\end{equation}
where $\gamma (t,s)=\gamma _0(s) + t \; \gamma _1(s)$ and  $y_{it}$ and  $W_{it}(\cdot)$
are the GDS and the basal ganglia spectrum for subject $i$ at time $t$, respectively. We
assume that $\epsilon_{it} \sim  N(0,\sigma_{\epsilon}^2)$ and $b_i$ is the
subject-specific random intercept distributed as $N(0,  \sigma_b^2)$. The estimates were
obtained as the BLUP from the mixed model formulation described in Section \ref{BLUP}
using $L_0=L_1=L_Q$.

The estimates of $\lambda$ (tuning parameter) associated with $\gamma_0(\cdot)$ and
$\gamma_1(\cdot)$ were $1.152$ and $2.242$, respectively and the estimates of
$\sigma_{\epsilon}^2$ and $\sigma_b^2$ were 0.0786 and 0.3332, respectively. The GDS
score, fitted values and residual plot are displayed in Figure~\ref{App2diagnostic} for
the purpose of model checking. The residuals do not show an obvious pattern indicating lack-of-fit of
 the proposed model.

Figure~\ref{App2} displays the estimates of $\gamma_0(\cdot)$ and $\gamma_1(\cdot)$ with
pointwise 95\% confidence bands. To aid interpretation, selected pure metabolite spectra
are displayed. These figures reveal that $\hat{\gamma}_0(\cdot)$ (the ``baseline" part of
the regression function) is different from zero at the locations where at least one of
the pure metabolites Cr, Glu, NAA, NAAG and Scyllo has a bump. Similarly, each non-zero
part of $\hat{\gamma}_1(\cdot)$  (the ``longitudinal" part of the regression function)
coincides with bump locations of one or more pure metabolite profiles of Cr, Glu, NAA,
GPC and Ins.

Pointwise confidence intervals for $\gamma_0(\cdot)$  and $\gamma_1(\cdot)$ contain the
$0$ line over large intervals. The estimated $\gamma_0$  is
significant in the region $s \in (0.4, 0.5)\cup (0.6,0.8))$ and estimated $\gamma_1$ is
significant in a region $s \in (0.5, 0.6)$. To be precise, peaks in both
$\hat{\gamma}_0(\cdot)$ and $\hat{\gamma}_1(\cdot)$ are significant at locations where at
least one of the pure metabolite profiles NAA or Glu have bumps. The observation of
negative `longitudinal' effect of NAA is worth commenting; it suggests that GDS increases
as NAA concentration decreases in basal ganglia, a finding consistent with several
studies in which a reduced concentration of NAA is seen to be associated with a decrease
in neuronal mass \citep{Christiansen1993, Lim1997, Soares2009}.

Finally, we considered other forms of $f(t)$, such as $exp(t)-1$ or $log(t+1)$. When
$\gamma(t,\cdot)=\gamma_0(t, \cdot) + [exp(t)-1] \gamma_1(t, \cdot)$ was compared with
$\gamma(t,\cdot)=\gamma_0(t, \cdot) + t \gamma_1(t, \cdot)$, the AIC increased to
$-394.56$ from $-395.23$. However, the estimation with $\gamma(t,\cdot)=\gamma_0(t,
\cdot) + log(t+1) \gamma_1(t, \cdot)$ did not show any non-zero regions for
$\gamma_1(\cdot)$, using a 95\% confidence band. This suggests that other time structures
in $\gamma(t,\cdot)$ may be useful, provided longitudinal observations are available for
longer time periods.

\section{Discussion}
We have proposed a novel estimation method for longitudinal functional regression and
derived some properties of the estimated coefficient function. A valuable contribution of
this framework is that it allows this estimate to vary with time as it extends the scope
of penalized regression to the realm of longitudinal data.  The approach may be viewed as
an extension of longitudinal mixed effects models, replacing scalar predictors by
functional predictors. Advantages of this framework include: estimating a time-dependent
regression function; the ability to incorporate structural information into the
estimation process; easy implementation through the linear mixed model equivalence.

The first simulation study of Section~\ref{Simulation1} illustrates the potential
advantage in exploiting an informed structured penalty, as compared to the more generic
smoothness or spline-based constraints. The simulation in Section~\ref{Simulation3}
suggests that coverage probabilities of the confidence bands for the true regression
function are close to the nominal level. However, for small sample sizes the naive
confidence bands do not seem to be sufficient and an alternative solution which takes
into account the estimation bias is needed. In the case when only partial information is
available the proposed method can be still useful, if we limit the relative contribution
of the ``informed" space and/or increase the sample size (see
Subsection~\ref{Simulation4}). In the absence of prior information, one may impose more
vaguely-defined constraints---such as identity penalties, smoothing penalties or
re-weighted projections onto empirical subspaces---to estimate the coefficient function.

Estimation in generalized ridge regression can be expressed in many forms.  Clearly, one
natural way to view this is via a Bayesian equivalence formulation \citep[see e.g.,
][]{BLUP} with the informative priors quantifying the available scientific knowledge. In
our formulation, the linear mixed model equivalence provides an easy computational
implementation as well as an automatic choice of the tuning parameters using REML
criterion. The GSVD provides algebraic insight and a convenient way to derive the bias
and variance expressions of the estimates.

A possible extension of this work is to incorporate multiple functional predictors. For
example, given two observed functional predictors $W^{(1)}_t(\cdot)$ and
$W^{(2)}_t(\cdot)$, consider two associated coefficient functions:
$\gamma^{(1)}(t,\cdot)$ and $\gamma^{(2)}(t,\cdot)$.  We can express $\gamma^{(1)}
(t,s)=\gamma^{(1)} _0(s)+f^{(1)}_1(t) \gamma ^{(1)}_1(s)+ \cdots +f^{(1)}_d(t)
\gamma^{(1)}_d(s)$ and $\gamma^{(2)}(t,s)=\gamma^{(2)} _0(s)+f^{(2)}_1(t)
+\gamma^{(2)}_1(s)+ \cdots +f^{(2)}_d(t) \gamma^{(2)}_d(s)$. Let $W^{(1)}$ and $W^{(2)}$
represent design matrices for the two functional predictors. Then we can estimate
$\gamma^{(1)}(t,\cdot)$ and $\gamma^{(2)}(t,\cdot)$ by finding the BLUP estimate of
$\gamma^{(1)}$  and $\gamma^{(2)}$ from  the  mixed model: \, $y=X\beta + W^{(1)}
\gamma^{(1)}+ W^{(2)}\gamma^{(2)} + Zb+\epsilon$.  The simplified formula for bias and
variance derived in Section~\ref{GSVD} still holds with an additional assumption that
$(W^{(1)})^{\top}V^{-1}W^{(2)}=0$.

As presented here, the method addresses models having a continuous scalar outcome, but
allowing for either binary or count responses is of interest. Indeed, an important
problem that arises in MRS data is that of understanding the neurocognitive impairment
status of HIV patients, defined as a binary variable, based on functional predictors
collected over time.  Estimation in these general settings appears to be possible with
the proposed framework.

\smallskip
\noindent {\large\bf Acknowledgment:}  The authors thank Dr.~B.~Navia who provided the
MRS data used as an example in the manuscript. Partial research support was provided by
the National Institutes of Health grants U01-MH083545 (JH), R01-CA126205 (TR) and
U01-CA086368 (TR).


\section{Connection with the GSVD} \label{GSVD}
We provide the derivation of a LongPEER estimate using the GSVD. This can be viewed as an
extension of the estimation discussed by \citet{PEER} in two ways: we allow for a general
covariance matrix $V$ (for $y$) and we extend the penalty operator to apply across
multiply-defined domains, $L_0,\ldots,L_D$.

After some algebra, the generalized ridge estimate in \eqref{RidgeEst} for $\gamma$ can
be expressed as
\begin{equation*}
\hat{\gamma}=-A_1X^{\top}V^{-1}y + A_2W^{\top}V^{-1}y
\end{equation*}
where
\begin{equation*}
\begin{aligned}
 A_1^{\top} &= (X^{\top}V^{-1}X)^{-1}X^{\top}V^{-1}W[W^{\top}V^{-1}W+L^{\top}L-W^{\top}V^{-1}X(X^{\top}V^{-1}X)^{-1}X^{\top}V^{-1}W]^{-1}\\
 A_2 &= W^{\top}V^{-1}W+L^{\top}L- W^{\top}V^{-1}X(X^{\top}V^{-1}X)^{-1}X^{\top}V^{-1}W
\end{aligned}
\end{equation*}
When $X=0$ (a situation without any scalar predictors) or $X^{\top}V^{-1}W =0$ the
generalized ridge estimation of $\gamma$ can be put into a PEER estimation framework in
terms of GS vectors, as discussed below.

With $X=0$ or $X^{\top}V^{-1}W =0$, the $\hat{\gamma}$ reduces to $[W^{\top}V^{-1}W+ L^{\top}L]^{-1}W^{\top}V^{-1}y$.
Moreover, in this case generalized ridge estimate of $\beta$ becomes
$[X^{\top}V^{-1}X]^{-1}X^{\top}V^{-1}y$. Now, if we transform $\tilde{W}:=V^{-1/2}W$ and
$\tilde{y}:=V^{-1/2}y$, we can rewrite $L$ as
\[
L=\lambda_0 \;\mbox{blockdiag}\left\{L_0, \frac{\lambda_1}{\lambda_0}L_1, \cdots, \frac{\lambda_D}{\lambda_0}L_D\right\} =\lambda_0 L^s
\]
Here, $L^s$ can be interpreted as a scaled $L$ where scaling is done for all the tuning
parameters associated with the `longitudinal' part of the regression function with
respect to the `baseline' tuning parameter.

Set $\tilde{p}=(D+1)p$, let $m$ denote the number of rows in $L$ and set
$c=\dim[\Null(L)]$.  Further, assume that $n_{\bullet}\leq m \leq \tilde{p} \leq
m+n_{\bullet}$ and the rank of the $(n_{\bullet}+m)\times \tilde{p}$ matrix
$[\tilde{W}^{\top}\; \;\; (L^s)^{\top}]^{\top}$ is $\tilde{p}$.  The following describes
the GSVD of the pair $(\tilde{W}, L^s)$: there exist orthogonal matrices $\mathcal{U}$
and $\mathcal{V}$, a nonsingular $\mathcal{G}$ and diagonal matrices $S$ and $M$ such
that
\[
\tilde{W}=\mathcal{U} \mathcal{S} \mathcal{G}^{-1}
\qquad \mathcal{S}=
\left[ 0 \;\; S\right]
\qquad S=\mbox{blockdiag}\{S_1,\;\; I_{\tilde{p}-m}\}
\]
\[
L^s=\mathcal{V} \mathcal{M} \mathcal{G}^{-1}
\qquad \mathcal{M}=\left[M \;\; 0\right]
\qquad M=\mbox{blockdiag}\{ I_{\tilde{p}-n_{\bullet}},\;\; M_1\}
\]
Submatrices $S_1$ and $M_1$  have $\ell=n_{\bullet}+m-\tilde{p}$ diagonal entries ordered
as
\[
\begin{array}{r}
0 < \sigma_1 \leq \sigma_2 \leq \cdots \leq \sigma_{\ell} < 1\\
0 > \mu_1 \geq \mu_2 \geq \cdots \geq \mu_{\ell} > 1\\
\end{array}
\qquad \mbox{where},
\qquad \sigma_k^2 +  \mu_k^2=1,
\qquad k = 1,\dots,\ell
\]
Here, the columns $\{g_k\}$ of $\mathcal{G}$ are the GS vectors determined by the GSVD of
the pair $(\tilde{W}, L^s)$. Denote the columns of $\mathcal{U}$ and $\mathcal{V}$ by
$u_{k}$ and $v_{k}$, respectively. Now, it can be shown that
$[W^{\top}V^{-1}W+L^{\top}L]^{-1}W^{\top}V^{-1} =[W^{\top}V^{-1}W+\lambda_0^2 (L^s)^{\top}L^s]^{-1}W^{\top}V^{-1}
=\mathcal{G}(\mathcal{S}^{\top}\mathcal{S}+\lambda_0^2\mathcal{M}^{\top}\mathcal{M})^{-1}\mathcal{G}^{\top}$ $\tilde{W}^{\top}V^{-1/2}$
and consequently, $\hat{\gamma}$ can be expressed as
\begin{equation*}
\hat{\gamma}
=\mathcal{G}(\mathcal{S}^{\top}\mathcal{S}+\lambda_0^2\mathcal{M}^{\top}\mathcal{M})^{-1}\mathcal{S}^{\top}\mathcal{U}^{\top}\tilde{y}
= \sum_{k=\tilde{p}-n_{\bullet}+1}^{\tilde{p}-c}{\frac{\sigma_k^2}{\sigma_k^2+\lambda_0^2\mu_k^2}\frac{1}{\sigma_k}u_{k}^{\top}\tilde{y}g_k}
+ \sum_{k=\tilde{p}-c+1}^{\tilde{p}}{u_{k}^{\top}\tilde{y}g_k} \nonumber
\end{equation*}
Further, the bias and variance can be expressed as
\begin{eqnarray}
Bias[\hat{\gamma}]=(I-W^{\#}W)\gamma& =\mathcal{G}(\mathcal{S}^{\top}\mathcal{S}+\lambda_0^2\mathcal{M}^{\top}\mathcal{M})^{-1}(\lambda_0^2\mathcal{M}^{\top}\mathcal{M})\mathcal{G}^{-1}\nonumber \\
&=\sum_{k=1}^{\tilde{p}-n_{\bullet}}{{g_k\tilde{g}_k^{\top}\gamma}}
       + \sum_{k=\tilde{p}-n_{\bullet}+1}^{\tilde{p}-c}{\frac{\lambda_0^2\mu_k^2}{\sigma_k^2+\lambda_0^2\mu_k^2}{g_k\tilde{g}_k^{\top}\gamma}} \nonumber
\end{eqnarray}
\begin{eqnarray}
Var[\hat{\gamma}]=W^{\#}V(W^{\#})^{\top}&= \mathcal{G}(\mathcal{S}^{\top}\mathcal{S}+\lambda_0^2\mathcal{M}^{\top}\mathcal{M})^{-1}\mathcal{S}^{\top}\mathcal{S}(\mathcal{S}^{\top}\mathcal{S}+\lambda_0^2\mathcal{M}^{\top}\mathcal{M})^{-1}\mathcal{G}^{\top}\nonumber \\
& =\sum_{k=\tilde{p}-n_{\bullet}+1}^{\tilde{p}-c}{\frac{\sigma_k^2}{(\sigma_k^2+\lambda_0^2\mu_k^2)^2}{g_kg_k^{\top}}}
 +      \sum_{k=\tilde{p}-c+1}^{\tilde{p}}{g_kg_k^{\top}} \nonumber
\end{eqnarray}
where, $W^{\#}=[W^{\top}V^{-1}W+ L^{\top}L]^{-1}W^{\top}V^{-1}$ and $\tilde{g}_k$ denotes
the $k$th column of
$\mathcal{G}^{-T}=(\mathcal{G}^{-1})^{\top}=(\mathcal{G}^{\top})^{-1}$. Further, we can
express bias as $[W^{\top}V^{-1}W+ L^{\top}L]^{-1} L^{\top}L\gamma$ which means
$\hat{\gamma}$ will be unbiased only when $\gamma \in \Null(L)$.

For estimates obtained using this technique, the bias and variance can be expressed in
terms of generalized singular vectors, provided the assumption of $X^{\top}V^{-1}W=0$
applies. In this case, one can show that $\hat{\beta}$ is simply the generalized least
squares estimate from the linear model $y=X\beta+\epsilon^*$, and $\hat{\gamma}$  is the
generalized ridge estimate from $y=W\gamma+\epsilon^*$ with penalty $L$. That is, $\beta$
is estimated as if $W\gamma$ were not present, and $\gamma$ is estimated as if $X\beta$
were not present.


{}


\vskip .65cm
\noindent
Department of Biostatistics
\vskip 2pt
\noindent
Indiana University Fairbanks School of Public Health
\vskip 2pt
\noindent
E-mail: mgkundu@iupui.edu
\vskip 2pt
\noindent
E-mail: harezlak@iupui.edu
\vskip .3cm


\end{document}